\documentclass[11pt,onecolumn]{IEEEtran}
\usepackage{mathpple}
\usepackage{times}
\usepackage{amsmath}  
\usepackage{amssymb}  
\usepackage{mathrsfs} 
\usepackage{theorem}  
\usepackage{cite}     
\usepackage{comment}  
\usepackage{upref}
\usepackage{amsfonts}
\usepackage{verbatim}
\usepackage[dvipsnames,usenames]{color}
\usepackage{enumerate}
\usepackage{graphicx}
\usepackage{subfigure}
\usepackage{latexsym}

\usepackage{psfrag}

\usepackage{color}
\usepackage{times,color}
\usepackage[usenames,dvipsnames,svgnames,table]{xcolor}

\newcommand{\code}{\ttfamily\bfseries}

\newcommand{\be}[1]{\begin{equation}\label{#1}}
\newcommand{\ee}{\end{equation}}

\newcommand{\hs}[1]{\hspace{-0.25ex}{#1}\hspace{-0.25ex}}

\newcommand{\bc}{\begin{center}}
\newcommand{\ec}{\end{center}}

\newcommand{\ceil}[1]{\lceil{#1}\rceil}

\newcommand{\qed}{\hfill$\Box$\\[1ex]}

\newcommand{\cA}{{\cal A}}

\newcommand{\cC}{{\cal C}}

\newcommand{\cE}{{\cal E}}

\newcommand{\cG}{{\cal G}}

\newcommand{\cN}{{\cal N}}
\newcommand{\cO}{{\cal O}}
\newcommand{\cP}{{\cal P}}
\newcommand{\cQ}{{\cal Q}}

\newcommand{\cS}{{\cal S}}

\newcommand{\cU}{{\cal U}}

\newcommand{\cX}{{\cal X}}

\newcommand{\bfa}{{\boldsymbol a}}

\newcommand{\bfc}{{\boldsymbol c}}

\newcommand{\bfe}{{\boldsymbol e}}

\newcommand{\bfu}{{\boldsymbol u}}

\newcommand{\bfx}{{\boldsymbol x}}

\renewcommand{\le}{\leqslant}
\renewcommand{\leq}{\leqslant}
\renewcommand{\ge}{\geqslant}
\renewcommand{\geq}{\geqslant}

\newcommand{\F}{\mathbb{F}}

\newcommand{\zero}{{\mathbf 0}}
\newcommand{\one}{{\mathbf 1}}

\newcommand{\Cref}[1]{Co\-rol\-la\-ry\,\ref{#1}}

\theoremstyle{plain} \theorembodyfont{\normalfont\slshape}

\newtheorem{thm}{Theorem$\!$}
\newenvironment{theorem}{\begin{thm}\hspace*{-1ex}{\bf.}}{\end{thm}}

\newtheorem{prop}[thm]{Proposition$\!$}

\newtheorem{lem}[thm]{Lemma$\!$}
\newenvironment{lemma}{\begin{lem}\hspace*{-1ex}{\bf.}}{\end{lem}}

\newtheorem{cor}[thm]{Corollary$\!$}

\newtheorem{const}[thm]{Construction$\!$}

\newtheorem{defi}[thm]{Definition$\!$}
\newenvironment{definition}{\begin{defi}\hspace*{-1ex}{\bf.}}{\end{defi}}

\theorembodyfont{\normalfont}

\newtheorem{exam}{Example$\!$}
\newenvironment{example}{\begin{exam}\hspace*{-1ex}{\bf .}}{\qed\end{exam}}

\newtheorem{remrk}{Remark$\!$}

\definecolor{Codecolor}{named}{White}

\newcommand{\Copen}{\mbox{\{\kern-5.50pt\{}}
\newcommand{\Cclose}{\mbox{\}\kern-5.50pt\}}}
\newcommand{\Cslash}{\mbox{$\backslash\kern-6.02pt\backslash$}}

\makeatletter

\DeclareRobustCommand{\sbinom}{\genfrac[]\z@{}}

\makeatother

\newcommand{\G}[2]{\sbinom{{#1}\kern-.05pt}{{#2}\kern-.05pt}}

\begin{document}

\title{\textbf{\huge{%
PIR with Low Storage Overhead:\\[0.50ex]
Coding instead of Replication\\[2.750ex]}}}

\author{
       \textbf{Arman Fazeli},\IEEEauthorrefmark{1}~~
       \textbf{Alexander Vardy},\IEEEauthorrefmark{1}~~
       \textbf{Eitan Yaakobi}\IEEEauthorrefmark{2}
\thanks{\IEEEauthorrefmark{1}%
Department of Electrical and Computer Engineering, 
Department of Computer Science and Engineering, and 
Department of Mathematics, 
University of California San Diego, La Jolla, CA\,92093,
USA (email: {\code afazelic@ucsd.edu}, {\code avardy@ucsd.edu}).\vspace{0.25ex}}
\thanks{\IEEEauthorrefmark{2}%
Department of Computer Science, 
Technion --- Israel Institute of Technology, 
Haifa, Israel (email: {\code yaakobi@cs.technion.ac.il}).}
}

\maketitle

\thispagestyle{empty}

\begin{abstract}
Private information retrieval (PIR) protocols allow a user to retrieve
a data item from a database without reveal\-ing any information about 
the identity of the item being retrieved. Specifically, in 
information-theoretic $k$-server~PIR, the database is replicated
among $k$ non-communicating servers, and each server learns nothing
about the item retrieved~by the user. The cost of
PIR protocols is usually measured in terms of their \emph{communication
complexity},~which is the total number of bits exchanged between 
the user and the servers. However, another important cost parameter
is the \emph{storage overhead}, which is the ratio between the total
number of bits stored on all the servers and the number of bits in
the database. Since single-server information-theoretic PIR is 
impossible, the storage overhead of all~existing PIR protocols
is at least $2$ (or $k$, in the case of $k$-server PIR).

In this work, we show that information-theoretic PIR can be
achieved with storage overhead arbitrarily close~to~the optimal
value of $1$, without sacrificing the communication complexity.
Specifically, we prove that \emph{all} known  
$k$-server PIR
protocols can be efficiently emulated, while preserving both
privacy and communication complexity~but~significantly reducing 
the storage overhead. To this end, we distribute the $n$ bits 
of the database among $s+r$ servers, each storing $n/s$ 
\emph{coded bits} (rather than replicas). 
Notably, our coding
scheme remains the same, regardless of the specific $k$-server
PIR protocol being emulated. 
For every fixed $k$, 
the resulting storage overhead $(s+r)/s$ approaches $1$ as
$s$~grows; explicitly we have $r \le k \sqrt{s}\bigl(1 + o(1)\bigr)$.
Moreover, in the special case $k = 2$,
the storage~overhead is\linebreak only \smash{$1 + \frac{1}{s}$}.
In order to achieve these results, we introduce~and study a new
kind of binary linear codes, called~here \emph{$k$-server PIR codes}.
We then show how such codes can be constructed from Steiner systems,
from one-step majority-logic decodable codes, from constant-weight 
codes, and from certain locally recoverable codes. 
We also establish~several 
bounds on the parameters of 
$k$-server PIR codes, and tabulate the results for all $s \le 32$
and $k \le 16$. Finally, we briefly discuss extensions 
of our results to nonbinary alphabets, to robust PIR, 
and to $t$-private PIR.
\end{abstract}

\vspace{5.00ex}
\section{Introduction} 
\label{sec:Intro}
\vspace{0.75ex}

\noindent
Private information retrieval protocols make it possible 
to retrieve a data item from a database without disclosing 
any information about the identity of the item being retrieved.
The notion of private information retrieval (PIR)
was first introduced by 
Chor, Goldreich, Kushilevitz, and Sudan in~\cite{CKGS95,CKGS98} 
and has attracted considerable attention since (see \cite{
BIK05,BIKR02,BIM00,CHY14,DG14,G04,WY05,Y10,Y08}
and references therein). The classic PIR model of~\cite{CKGS98},
which we adopt~in~this paper, views the database as a binary 
string $\bfx = (x_1,\ldots,x_n) \in \{0,1\}^n$ and assumes that
the user wishes~to~retrieve\linebreak 
a single bit $x_i$ without revealing
any information about the index $i$. A naive solution for the 
user (hereinafter, often called Alice) is to download the entire
database $\bfx$. It is shown in~\cite{CKGS98} that in the case
of a single database stored on a single server, this solution
is essentially the best possible: any PIR protocol will require
$\Omega(n)$ bits~of~communication between the user and the server.
In order to achieve sublinear communication complexity, 
Chor, Goldreich, Kushilevitz, and Sudan~\cite{CKGS98} 
proposed \emph{replicating the database} on several servers
that do not communicate with each other. They showed that
having two replicas 
makes it possible to reduce the communication cost to $O\bigl(n^{1/3}\bigr)$, 
while $k \ge 3$ servers can achieve
\smash{$O\bigl((k^2\log k)n^{1/k}\bigr)$} communication complexity.

\looseness=-1
Following the seminal work of~\cite{CKGS98}, the communication complexity
of $k$-server PIR has been further reduced in a~series of groundbreaking
papers. Ambainis~\cite{Ambainis79} generalized the methods of~\cite{CKGS98} 
to obtain a communication cost of \smash{$O\bigl(n^{1/(2k-1)}\bigr)$} 
for all $k \ge 2$. 
This result remained the best known for a while until the
\smash{$O\bigl(n^{1/(2k-1)}\bigr)$}-complexity~barrier
was finally broken in \cite{BIKR02}.
Five years later, came the remarkable work of
Yekhanin~\cite{Y08} who constructed~a~$3$-server 
PIR scheme with subpolynomial communication
cost, assuming the infinitude of Mersenne primes.
Shortly thereafter, Efremenko~\cite{Efremenko} gave an unconditional 
$k$-server PIR scheme with subpolynomial complexity for all $k \ge 3$.
The recent paper of Dvir and Gopi \cite{DG14} shows how to
achieve the same complexity as in~\cite{Efremenko} 
with only two servers.

\looseness=-1
All this work follows the orginal idea, first proposed in~\cite{CKGS98},
of replicating the database in order to reduce the~communication cost.
However, this approach neglects another  
cost parameter: 
the \emph{storage overhead}, defined~as the ratio between the total
number of bits stored on all the servers and the number of bits in
the database. Clearly, the storage overhead of the PIR protocols
discussed above is $k\ge2$. 
If the database is very large, the necessity to store several
replicas of it could be untenable for some applications.
Thus, in this paper, we consider the
following question. Can one achieve PIR with low communication cost but
\emph{without doubling (or worse) the number of bits we need to store}?~~

This question has been settled in the affirmative in~\cite{KO}
for the case where one is willing to replace information-theoretic
guarantees of privacy by computational guarantees. Such computational
PIR is by now well studied ---~see \cite{KO,Gasarch} for more information.
However, in this paper, we consider only information-theoretic PIR,
which provides the strongest form of privacy. That is, even 
computationally unbounded servers should not gain any 
information~from the user queries. Somewhat 
surprisingly, despite the impossibility proof of~\cite{CKGS98}, the 
answer to our question turns out to~be affirmative
also in the case of information-theoretic PIR. Our
results do not contradict~\cite{CKGS98}. 
To achieve~in\-formation-theoretic privacy, one does need at least two
non-communicating servers. However, these servers
do not have to hold the entire database, they can 
store only parts of it. We show that if these parts are judiciously
encoded, rather than replicated, the overall storage overhead
can be reduced.

\vspace{1.50ex}
\subsection{Our Contributions}\label{subsec:our_contribution}

We show that all known $k$-server information-theoretic PIR protocols
can be efficiently emulated, while preserving their 
privacy and communication-complexity guarantees (up to a constant),
but significantly reducing the storage overhead.
In fact, for any fixed $k$ and any $\epsilon > 0$,
we can reduce the storage overhead to under $1 + \epsilon$.

In order to achieve these results, we first partition the database
into $s$ parts and distribute these parts among~non-communicating
servers, so that every server stores $n/s$ bits. Why do we  
partition the database in this manner?  The main reason is that
such partition is necessary to reduce the storage overhead. If every
server has to store all~$n$~bits of the database, then the storage
overhead cannot be reduced beyond $k \ge 2$. However, in practice,
there may be other compelling reasons. For example, 
the database may be simply too large to fit in a single server, 
or it may need to be stored in a distributed manner for security purposes.
We observe that the number of parts $s$ need not be very large. 
With $s=2$ parts, we can already achieve significant savings
in storage overhead. With $s=16$ parts,~we get a storage overhead
of $1.06$ (for $2$-server PIR protocols).

Given a partition of the database into $s$ parts,
our construction consists of two main
ingredients: 1) an existing $k$-server PIR protocol in which the
servers' responses are a \emph{linear function} of the database bits, 
and 2) a binary~linear code, which we call a \emph{$k$-server PIR code}, 
with a special property to be defined shortly.
We note that the~first~re\-quirement is very easy to satisfy:
\emph{all} the existing PIR protocols known (to us) are linear in this
fashion. Thus our primary focus in this paper is on the construction 
of $k$-server PIR codes.

The defining property of a $k$-server PIR code is this: for
every message bit $u_i$, there exist $k$ disjoint sets of coded
bits from which $u_i$ can be uniquely recovered (see 
Section\,\ref{sec:coded_PIR} for a formal definition).
Although this property is reminiscent of locally
recoverable codes, recently introduced in~\cite{GHSY12}, 
there are important differences. In locally~recoverable
codes, we wish to guarantee that every message bit $u_i$
can be recovered from a \emph{small set} of coded bits,
and only one such recovery set is needed. Here, we wish
to have \emph{many disjoint recovery sets} for every 
message~bit, and we do not care about their size.
To the best of our knowledge, codes with this property
have not been previously studied, and they may be of
independent interest.

In this paper, we show how $k$-server PIR codes can 
be constructed from Steiner systems,
from one-step majority-logic decodable codes,
and from constant-weight codes. We give an optimal
construction of such codes for the case where the 
number of parts $s$ is small. We also establish several 
bounds on the parameters of general
$k$-server PIR codes, and tabulate these parameters for all $s \le 32$
and $k \le 16$. Finally, we briefly discuss extensions 
of our results to nonbinary alphabets, to robust PIR, 
and to $t$-private PIR.

\subsection{Related work}\label{subsec:related_work}

There are several previous works which construct coded schemes for the purpose of fast or private retrieval. The first work we know of for the purpose of coded private retrieval is the recent work by Shah et al.~\cite{SRR14}. The authors showed how to encode files in multiple servers with very low communication complexity. However, their constructions require an exponentially large number of servers which may depend on the number of files or their size.
In another recent work~\cite{CHY14}, Chan et al. studied the tradeoff between storage overhead and communication complexity, though only for setups in which the size of each file is relatively large.
A similar approach to ours was studied by Augot et al.~\cite{ALS14}, where the authors also partitioned the database into several parts in order to avoid repetition and thereby reduce the storage overhead. However, their construction works only for the PIR scheme using the multiplicity codes by Kopparty et al.~\cite{KSY11} and they didn't encode the parts of the database as we study in this work.

Batch codes~\cite{IKOS04} are another method to store coded data in a distributed storage for the purpose of fast retrieval of multiple bits. Under this setup, the database is encoded into an $m$-tuple of strings, called \emph{buckets}, such that each batch of $k$ bits from the database can be recovered by reading at most some predetermined $t$ bits from each bucket. 
They are also useful in trading the storage overhead in exchange for load-balancing or lowering the computational complexity in private information retrieval. 
Another recent work on batch codes was recently studied in~\cite{DGRS14}.

\subsection{Organization}

The rest of this paper is organized as follows. In Section~\ref{sec:prelim}, we formally define the PIR schemes studied in the paper, namely the conventional PIR protocol and coded PIR protocols. In Section~\ref{sec:coded_PIR} we present our general construction of coded PIR protocols and define the requirements on a $k$-server PIR codes that are used in this protocol. Section~\ref{sec:cons} studies several constructions of $k$-server PIR codes. In Section~\ref{sec:small_s_k}, we study the storage overhead of $k$-server PIR codes when the values of $s$ and $k$ are small, and in Section~\ref{sec:asymp_behavior} we study the asymptotic behavior when either $s$ or $k$ is large. Finally, Section~\ref{sec:conc} concludes the paper.

\section{Definitions and Preliminaries}\label{sec:prelim}

In this section we formally define the PIR protocols we study in the paper. A linear code over $GF(q)$ of length $n$ and dimension $k$ will be denoted by $[n,k]_q$ or by $[n,k,d]_q$ where $d$ specifies the minimum distance of the code. In case the code is binary we will omit the field notation. For a positive integer $n$ the notation $[n]$ will refer to the set $\{1,\ldots,n\}$. We denote by $\bfe_i$ the vector with 1 on its $i$-th position and 0 elsewhere. Let us revisit and rephrase the formal definition of a PIR scheme, based upon the definitions taken from~\cite{BIKR02} and~\cite{G04}.

\begin{definition}\label{def:PIR}
A \textbf{$k$-server PIR scheme} consists of the following:
\begin{enumerate}
\item $k$ servers $\cS_1,\ldots,\cS_k$, each stores a length-$n$ database $\bfx$,
\item A user (Alice) $\cU$ who wants to retrieve $x_i$, for $i\in [n]$, without revealing $i$.
\end{enumerate}
A \textbf{$k$-server PIR protocol} is a triplet of algorithms $\cP = (\cQ, \cA, \cC)$ consisting of the following steps:
\begin{enumerate}
\item\label{step 1} Alice flips coins and based on the flip coins and $i$ invokes $\cQ(k,n;i)$ to generate a randomized $k$-tuple of queries $(q_1,\ldots, q_k)$, of some predetermined fixed length. For $j\in [k]$, the query $q_j$ will be also denoted by $\cQ_j(k,n;i)$. 
\item For $j\in [k]$, she sends the query $q_j$ to the $j$-th server $\cS_j$. 
\item The $j$-th server, for $j\in [k]$, responds with an answer $a_j=\cA(k,j,\bfx,q_j)$ of some fixed length.
\item Finally, Alice computes its output by applying the reconstruction algorithm $\cC(k,n;i,a_1,\ldots,a_k)$.
\end{enumerate}
The protocol should satisfy the following requirements:
\begin{itemize}
\item \textbf{Privacy} - Each server learns no information about $i$. Formally, for any $k,n,i_1,i_2 \in [n]$, and a server $j\in[k]$, the distributions $\cQ_j(k,n;i_1)$ and $\cQ_j(k,n;i_2)$ are identical, where the distribution is over the coins flip in Step~\ref{step 1} of the PIR protocol.
\item \textbf{Correctness} - For each $k,n$ and $\bfx\in\{0,1\}^n$ and $i\in[n]$, the user deterministically outputs the correct value of $x_i$, that is $\cC(k,n;i,a_1,\ldots,a_k) = x_i$. 
\end{itemize}
\end{definition}
We follow the common figure of merit to evaluate the system storage efficiency according by its overhead~\cite{SRR14,IKOS04}. Namely, the \emph{storage overhead} of the system is the ratio between the total number of bits stored in the system and the number of information bits. For example, the storage overhead of a $k$-server PIR scheme is $k$.

Another special property of PIR protocols which will be used in our constructions is \emph{linearity}. This property is formally defined as follows.

\begin{definition}\label{ded:linear_PIR}
A $k$-server PIR protocol $\cP = (\cQ, \cA, \cC)$ is said to be a \textbf{linear PIR protocol} if for every $n$, $j\in[k]$, $\bfx_1,\bfx_2\in\{0,1\}^n$, and a query $q_j$, the following property holds\vspace{-1ex}
$$  \cA(k,j,\bfx_1+\bfx_2,q_j) =  \cA(k,j,\bfx_1,q_j) + \cA(k,j,\bfx_2,q_j).$$
\end{definition}
Many, if not all existing PIR protocols, satisfy this linearity property, see for example~\cite{BIK05, BIKR02, BIM00,CKGS98, DG14, WY05, Y10, Y08}.
Lastly, we assume that the algorithm $\cA$ is public knowledge in the sense that every server can compute the response $\cA(k,j,\bfx,q)$ for any $j\in[k]$, database $\bfx$, and query $q$.

Before formally defining the coded version of a PIR scheme, we demonstrate the main ideas in the next example.

\begin{example}
Consider the following 2-server PIR scheme where each server stores an $n$-bit database $\bfx$ and Alice wants to read the $i$-th bit $x_i$, for some $i\in [n]$. Alice chooses uniformly at random a vector $\bfa\in\{0,1\}^n$. The first server receives the query $\bfa$ and responds with an answer of the bit $\bfa \cdot \bfx$. The second server receives the query $(\bfa+\bfe_i)$ and responds with an answer of the bit $(\bfa +\bfe_i)\cdot \bfx$; see Fig.~\ref{fig:ex1}. Alice receives these two bits and their sum gives the $i$-th bit $x_i$, since
$$\bfa \cdot \bfx + (\bfa +\bfe_i)\cdot \bfx = \bfa \cdot \bfx + \bfa \cdot \bfx + \bfe_i \cdot \bfx = x_i.$$

If the servers do not communicate with each other then since the vector $\bfa$ is chosen uniformly at random, the value of $i$ remains private. Moreover, the servers' responses are linear functions of the stored data and thus the protocol is a linear PIR protocol. Alice had to transmit $2n$ bits and $2$ bits were received, so a total of $2n+2$ bits were communicated. The storage overhead of this scheme is 2 and note also that if one of the servers fails then it is possible to retrieve the database $\bfx$ from the other surviving server.

Now, assume that the database $\bfx$ is partitioned into two equal parts of $n/2$ bits each, $\bfx_1$ and $\bfx_2$, where $\bfx_1=(x_1,\ldots, x_{n/2})$, and $\bfx_2=(x_{n/2+1},\ldots,x_n)$. The database is stored in three servers. The first server stores $\bfx_1$, the second stores $\bfx_2$, and the third one is a parity server which stores $\bfx_1+\bfx_2$. If Alice wants to read the $i$-th bit where $i\in [n/2]$, she first chooses uniformly at random a vector $\bfa\in\{0,1\}^{n/2}$. The first server receives the query $\bfa$ and responds with the bit $\bfa\cdot \bfx_1$. The second server receives the query $\bfa +\bfe_i$ and responds with the bit $(\bfa +\bfe_i)\cdot \bfx_2$, and the third server receives the query $\bfa +\bfe_i$ and responds with the bit $(\bfa +\bfe_i)\cdot (\bfx_1+\bfx_2)$. Alice receives those three bits and calculates the bit $x_i$ according to the sum
\begin{align*}
\bfa\cdot \bfx_1 + (\bfa +\bfe_i)\cdot \bfx_2 + (\bfa +\bfe_i)\cdot (\bfx_1+\bfx_2) = \bfa\cdot \bfx_1 + (\bfa +\bfe_i)\cdot \bfx_1 = x_i.
\end{align*}

It is clear that both schemes keep the privacy of $i$. In the first scheme, the number of communicated bits is $2n+2$, while in the coded scheme it is $3n/2+3$. The storage overhead was improved from 2 to 3/2, and both schemes can tolerate a single server failure. However, we note that the coded scheme requires one more server.
\begin{figure}[ht!]
\centering
\includegraphics[width=90mm]{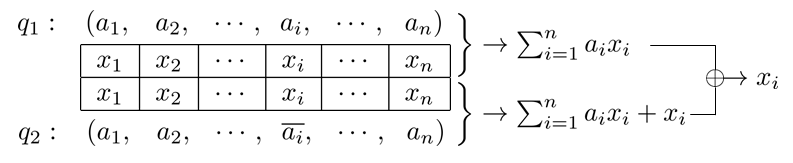}
\caption{{Alice sends $q_1=\bfa$ and $q_2=\bfa+\bfe_i$ to the servers. The servers respond with $\bfa \cdot \bfx$ and $(\bfa + \bfe_i) \cdot \bfx$ and Alice recovers $x_i$ as their sum. The value of $i$ remains private as the vector $\bfa$ is chosen uniformly at random.}}\label{fig:ex1}
\end{figure}
\end{example}

One may claim that the improvement in the last example is the result of using three instead of two servers. This is indeed correct, however, assume that each server can store only $n/2$ bits. Then, the database $\bfx$ will have to be stored over two servers and each of them would have to be replicated, resulting with a total of four servers instead of three. Furthermore, the number of communicated bits would still remain the same, $2n+2$. Thus, under this constraint, we can claim that we improved both the number of servers as well as the number of communicated bits.

We are now ready to extend the definition of PIR scheme to its coded version.
\begin{definition}\label{def:coded_PIR}
An \textbf{$(m,s)$-server coded PIR scheme} consists of the following:
\begin{enumerate}
\item A length-$n$ database $\bfx$ which is partitioned into $s$ parts $\bfx_1,\ldots,\bfx_s$, each of length $n/s$.
\item $m$ servers $\cS_1,\ldots,\cS_{m}$, where for $j\in [m]$ the coded data $\bfc_j$, stored in the $j$-th server, is a function of $\bfx_1,\ldots,\bfx_s$.
\item A user (Alice) $\cU$ who wants to retrieve the $i$-th bit from the database $\bfx$, without revealing $i$.
\end{enumerate}
An \textbf{$(m,s)$-server coded PIR protocol} is a triplet of algorithms $\cP^* = (\cQ^*, \cA^*, \cC^*)$ consisting of the following steps:
\begin{enumerate}
\item Alice flips coins such that based on the flip coins and $i$, she invokes $\cQ^*(m,s,n;i)$ to generate a randomized $m$-tuple of queries $(q_1,\ldots,q_{m})$ of predetermined fixed length.
\item For $j\in[m]$, she sends the query $q_j$ to the $j$-th server $\cS_j$.
\item The $j$-th server, for $j\in [m]$, responds with an answer $a_j=\cA^*(m,s,j,\bfc_j,q_j)$.
\item Finally, Alice computes its output by applying the reconstruction algorithm $\cC^*(m,s,n;i,a_1,\ldots,a_{m})$.
\end{enumerate}
The protocol should satisfy the privacy and correctness properties as stated in Definition~\ref{def:PIR}.
\end{definition}
The next section discusses the construction of coded PIR schemes based upon existing linear PIR protocols.

\section{Coded PIR Schemes}\label{sec:coded_PIR}
In this section we will give a general method to construct coded PIR schemes. A key point in the construction of coded PIR protocols is to use existing PIR protocols and emulate them in the coded setup. We first give a detailed example that demonstrates the main principles of the construction.

\begin{example}\label{ex:coded_PIR}
Assume there exists a 3-server linear PIR protocol $\cP(\cQ,\cA,\cC)$ and a length-$n$ database $\bfx$. Assume also that each server can store at most $n/4$ bits. If one wishes to invoke the PIR protocol $\cP(\cQ,\cA,\cC)$, then first the database $\bfx$ will be partitioned into four parts $\bfx_1,\bfx_2,\bfx_3,\bfx_4$. Thus, each of the four parts will be stored in three servers so it is possible to invoke the 3-server PIR protocol. This results with 12 servers, each stores $n/4$ bits, and thus the storage overhead is $3$. We will show how it is possible to accomplish the same task with storage overhead 2, that is only 8 instead of 12 servers. Namely, we construct an $(8,4)$-server coded PIR protocol $\cP^*(\cQ^*,\cA^*,\cC^*)$.

We use a similar partition of the database $\bfx$ into four parts $\bfx_1,\bfx_2,\bfx_3,\bfx_4$ and encode them into 8 servers as follows.
The $j$-th server for $j\in[8]$ stores the coded data $\bfc_j$ as follows:
$$\hspace{-1.5ex}\begin{array}{llll}
\bfc_1=\bfx_1, 			& \bfc_2=\bfx_2, 		& \bfc_3=\bfx_3, 		& \bfc_4=\bfx_4, \\
\bfc_5=\bfx_1+\bfx_2, 	& \bfc_6=\bfx_2+\bfx_3, & \bfc_7=\bfx_3+\bfx_4, & \bfc_8=\bfx_4+\bfx_1.
\end{array}$$
In a matrix form notation, these equations can be written in the following way
$$(\bfc_1,\ldots,\bfc_8) = (\bfx_1,\bfx_2,\bfx_3,\bfx_4) \cdot \left( \begin{array}{cccccccc}
1 & 0 & 0 & 0 & 1 & 0 & 0  & 1 \\
0 & 1 & 0 & 0 & 1 & 1 & 0  & 0 \\
0 & 0 & 1 & 0 & 0 & 1 & 1  & 0 \\
0 & 0 & 0 & 1 & 0 & 0 & 1  & 1
\end{array} \right).$$
Thus, we encode using an $[8,4,3]$ linear code where the last matrix is its generator matrix in a systematic form. 

Assume Alice seeks to read the $i$-th bit from the first part of the database $\bfx$, i.e. the bit $x_{1,i}$, or $x_i$ where $i\in[n/4]$. She first invokes algorithm $\cQ$ to receive the following three queries,
 $$\cQ(3,n/4;i) = (q_1,q_2,q_3).$$
Then, she assigns the 8 queries of the algorithm $\cQ^*$ to the 8 servers as follows
$$\cQ^*(8,4,n;i) = (q_1,q_2,q_3,q_3,q_2,q_2,q_3,q_3).$$
Next, she sends these queries to the servers, which respond with the following answers as listed in Table~\ref{PIR 1st}.
\begin{table}[htdp]
\caption{PIR Protocol for retrieving from the first server}
\begin{center}
\begin{tabular}{|c|c|c|}
\hline
Server & Query & Response \\ \hline \hline
1 & $q_1$ & $a_1=\cA^*(8,4,1,\bfc_1,q_1) = \cA(3,1,\bfx_1,q_1)$\\ \hline
2 & $q_2$ & $a_2=\cA^*(8,4,2,\bfc_2,q_2) = \cA(3,2,\bfx_2,q_2)$\\ \hline
4 & $q_3$ & $a_4=\cA^*(8,4,4,\bfc_4,q_3) = \cA(3,3,\bfx_4,q_3)$\\ \hline
5 & $q_2$ & $a_5=\cA^*(8,4,5,\bfc_5,q_2) = \cA(3,2,\bfc_5=\bfx_1+\bfx_2,q_2)$\\ \hline
8 & $q_3$ & $a_8=\cA^*(8,4,8,\bfc_8,q_3) = \cA(3,3,\bfc_8=\bfx_4+\bfx_1,q_3)$\\ \hline
\end{tabular}
\end{center}
\label{PIR 1st}\vspace{-4ex}
\end{table}

Due to the linearity property of the protocol $\cP$, Alice can calculate the following information from the second and fifth servers
\begin{align*}
a_2' &= a_2+a_5 = \cA(3,2,\bfx_2,q_2) + \cA(3,2,\bfc_5=\bfx_1+\bfx_2,q_2)&\\
 & = \cA(3,2,\bfx_1,q_2), &
\end{align*}
and from the fourth and eighth servers
\begin{align*}
a_3' &= a_4+a_8 = \cA(3,3,\bfx_4,q_3) + \cA(3,3,\bfc_8=\bfx_4+\bfx_1,q_3)&\\
 & = \cA(3,3,\bfx_1,q_3). &
\end{align*}
She also assigns $a_1'=a_1$.
Finally, Alice retrieves the value of $x_{1,i}$ by applying the reconstruction algorithm
\begin{align*}
& \cC^*(4,4,n;i,a_1,\ldots,a_{8}) = \cC(3,n/4;i,a_1',a_2',a_3') & \\
& = \cC(3,n/4;i,\cA(3,1,\bfx_1,q_1),\cA(3,2,\bfx_1,q_2),\cA(3,3,\bfx_1,q_3)) &\\
& = x_{1,i}=x_i. &\vspace{-2ex}
\end{align*}
Now, assume that Alice wants to retrieve the $i$-th bit from the second server, $x_{2,i}$, or $x_{n/2+i}$ for $i\in [n/2]$. As in the first case she invokes the algorithm $\cQ$ to receive
$$\cQ^*(4,4,n;i) = (q_2,q_1,q_3,q_3,q_2,q_3,q_3,q_3),$$
where $q_1,q_2,q_3$ are calculated according to $\cQ(3,n/4;i) = (q_1,q_2,q_3)$.
However, the queries to the servers will be slightly different, as summarized in Table~\ref{PIR 2nd}.
\begin{table}[htdp]
\caption{PIR Protocol for retrieving from the second server}
\begin{center}
\begin{tabular}{|c|c|c|}
\hline
Server & Query & Response \\ \hline \hline
1 & $q_2$ & $a_1=\cA^*(8,4,1,\bfc_1,q_2) = \cA(3,2,\bfx_1,q_2)$\\ \hline
2 & $q_1$ & $a_2=\cA^*(8,4,2,\bfc_2,q_1) = \cA(3,1,\bfx_2,q_1)$\\ \hline
3 & $q_3$ & $a_3=\cA^*(8,4,3,\bfc_3,q_3) = \cA(3,3,\bfx_3,q_3)$\\ \hline
5 & $q_2$ & $a_5=\cA^*(8,4,5,\bfc_5,q_2) =\cA(3,2,\bfc_5=\bfx_1+\bfx_2,q_2)$\\ \hline
6 & $q_3$ & $a_6=\cA^*(8,4,6,\bfc_6,q_3) =\cA(3,3,\bfc_6=\bfx_2+\bfx_3,q_3)$\\ \hline
\end{tabular}
\end{center}
\label{PIR 2nd}
\end{table}

Her next step is to calculate the following
\begin{align*}
a_2' &= a_1+a_5 = \cA(3,2,\bfx_1,q_2) + \cA(3,2,\bfc_5=\bfx_1+\bfx_2,q_2)&\\
 & = \cA(3,2,\bfx_2,q_2), &  \\
 a_3' &= a_3+a_6 = \cA(3,3,\bfx_3,q_3) + \cA(3,3,\bfc_6=\bfx_2+\bfx_3,q_3)&\\
 & = \cA(3,3,\bfx_2,q_3), &
\end{align*}
and to assign $a_1'=a_2$. Finally, Alice retrieves the value of $x_{2,i}$ by applying the reconstruction algorithm
\begin{align*}
& \cC^*(4,4,n;i,a_1,\ldots,a_{8}) = \cC(3,n/4;i,a_1',a_2',a_3') & \\
& = \cC(3,n/4;i,\cA(3,1,\bfx_2,q_1),\cA(3,2,\bfx_2,q_2),\cA(3,3,\bfx_2,q_3)) &\\
& = x_{2,i} = x_{n/2+i}. &
\end{align*}
\end{example}

In this example, we have not specified the queries to the other servers simply because they do not matter for the reconstruction. Thus we can assign any query to preserve the privacy property. However, we note that the requests to the servers might differ depending on the part of the database from which Alice wants to read the bit, and the different requests might reveal the identity of the data (in case the algorithm $\cA$ does not depend on $j$ then clearly this is not a problem). A simple solution for this is to ask every server to return all possible outputs such that it simulates each of the possible $k$ requests. For example, if in the first scenario Alice assigns the query $q_1$ to the first server, then the server's output consists of three parts:
\begin{align*}
\cA^*(8,4,1,\bfx_1&,q_1) =  (\cA(3,1,\bfx_1,q_1), \cA(3,2,\bfx_1,q_1), \cA(3,3,\bfx_1,q_1)).
\end{align*}
That way, Alice can choose the information required in order to compute the bit she wants to retrieve, and the server cannot deduce which part of the database the bit is read from. Yet another solution, which improves the download complexity, will be given as part of the proof of Theorem~\ref{thm:coded_PIR} below.

As we saw in Example~\ref{ex:coded_PIR}, there are two important ingredients in the construction of coded PIR protocols:
\begin{enumerate}
\item A $k$-server linear PIR protocol.
\item An $[m,s]$ linear code with special properties which are next specified.
\end{enumerate}
\begin{definition}\label{def:PIR code}
We say that an $s \times m$ binary matrix $G$ has property $A_k$
if for every $i \in [s]$, there exist $k$ disjoint subsets of
columns of $G$ that add up to the vector of weight one,
with the single $1$ in position $i$.
A binary linear $[m,s]$ code $C$ will be called a \textbf{$k$-server PIR code}
if there exists a generator matrix $G$ for $C$ with property $A_k$.
Equivalently, let $\bfc = \bfu G$ be the encoding of a message
$\bfu \in \{0,1\}^s$. Then $C$ is a {$k$-server PIR code}
if for every $i\in [s]$, there exist $k$ disjoint sets
$R_1,\ldots,R_k\subseteq [m]$ such that
$$ u_i \ = \ \sum_{j \in R_1} c_j \ = \ \cdots \ = \ \sum_{j \in R_k} c_j.$$
\end{definition}

The construction of $k$-server PIR codes will be deferred to Section~\ref{sec:cons}. In particular, we will be interested in finding, for given $s$ and $k$ the optimal $m$ such that an $[m,s]$ $k$-server PIR code exists, and the optimal value of $m$ will be denoted by $A(s,k)$. 
In terms of the minimum distance, let us briefly note that the minimum distance of a $k$-server PIR code is at least $k$.

We finish this section with the next theorem which provides the general result for the construction of coded PIR protocols. In order to analyze the communication complexity, we denote the number of bits uploaded, downloaded of a  $k$-server linear PIR protocol $\cP$, by $U(\cP;n,k)$, $D(\cP;n,k)$, respectively. For an $(m,s)$-server coded PIR protocol $\cP^*$, $U^*(\cP;n,m,s)$, $D^*(\cP;n,m,s)$ are defined similarly.

\begin{theorem}\label{thm:coded_PIR}
If there exists an $[m,s]$ $k$-server PIR code $\cC$ and a $k$-server linear PIR protocol $\cP$ then there exists an $(m,s)$-server coded PIR protocol $\cP^*$. Furthermore,
\begin{align*}
& U^*(\cP^*;n,m,s) = m\cdot U(\cP;n/s,k),  \\
& D^*(\cP^*;n,m,s) = m\cdot D(\cP;n/s,k).
\end{align*}
\end{theorem}
\begin{IEEEproof}
The database $\bfx$ is partitioned into $s$ parts $\bfx_1,\ldots, \bfx_s$. Let $G$ be a generator matrix of the code $\cC$. Then, the data stored in the $m$ servers is encoded according to
$$(\bfc_1,\ldots,\bfc_{m})= (\bfx_1,\ldots, \bfx_s) \cdot G.$$ Let $\cP(\cQ,\cA,\cC)$ be a $k$-server linear PIR protocol and we will show how to construct an $(m,s)$-server coded PIR protocol.

Assume Alice wants to read the $i$-th bit, $i\in [n/s]$, from the $\ell$-th server, $x_{\ell,i}$. First she invokes the algorithm $\cQ$ to receive $k$ queries $$\cQ(k,n/s;i) = (q_1,\ldots,q_k).$$
According to the $k$-server PIR code $\cC$, there exist $k$ mutually disjoint sets $R_{\ell,1},\ldots,R_{\ell,k}\subseteq [m]$ such that for all $ j\in [k]$, $\bfx_\ell$ is a linear function of the data stored in the servers belonging to the set $R_{\ell,j}$, that is,
$$\bfx_\ell = \sum_{h\in R_{\ell,j}}\bfc_h.$$

Alice assigns the output of the algorithm $\cQ^*$ to be
$$\cQ^*(m,s,n;i) = (q_1^*,\ldots,q_{m}^*),$$
where for all $j\in[k]$ and $h\in R_{\ell,j}$, $q_h^* = q_j$. The other queries $q_h^*$ where $h\notin \cup_{j\in[k]}R_{\ell,j}$ can be assigned arbitrarily. Then Alice sends the query $q_h^*$ to the $h$-th server, $h\in [m]$, and receives the answer
$$a_h^* = \cA^*(s,r,h,\bfc_h,q_h^*) = (\cA(k,1,\bfc_h,q_h^*),\ldots, \cA(k,k,\bfc_h,q_h^*)).$$ From these answers she takes only the parts which are required to invoke the algorithm $\cC$, and are determined by
$$a_h = \cA(k,j,\bfc_h,q_h^*),$$ for $h\in R_{\ell,j}$. Then, she assigns for $j\in [k]$,
$$a_j' = \sum_{h\in R_{\ell,j}} a_h,$$ and finally Alice retrieves the value of $x_{\ell,i}$ by applying the reconstruction algorithm
$$\cC(k,n/s;i,a_1',a_2',\ldots,a_k') = x_{\ell,i}.$$

The correctness of the last step results from the linearity of the PIR protocol $\cP$, since for all $j\in [k]$,
\begin{align*}
a_j' & = \sum_{h\in R_{\ell,j}} a_h = \sum_{h\in R_{\ell,j}} \cA(k,j,\bfc_h,q_h^*)  \\
& = \cA(k,j,\sum_{h\in R_{\ell,j}}\bfc_h,q_h^*) = \cA(k,j,\bfx_\ell,q_h^*) =  \cA(k,j,\bfx_\ell,q_j). 
\end{align*}
Therefore,
\begin{align*}
& \cC(k,n/s;i,a_1',a_2',\ldots,a_k')  \\
& = \cC(k,n/s;i,\cA(k,1,\bfx_\ell,q_j),\ldots,\cA(k,k,\bfx_\ell,q_j)) = x_{\ell,i}. 
\end{align*}

Let us add the following modification to this proof, in order to keep the privacy of the part in which Alice reads a bit. When Alice invokes the algorithm $\cQ$ and receives the $k$ queries $\cQ(k,n/s;i)=(q_1,\ldots, q_k)$, she also flips a coin to choose uniformly at random a permutation $\sigma$ of the elements in $[k]$ and assigns for all $j\in[k]$, $\hat{q}_j = q_{\sigma(j)}$. She continues with the algorithm to set for each $j\in[k]$ and $h\in R_{\ell,j}$, $q_{h}^*=q_{\sigma(j)}$. Then, the $h$-th server responds with the answer
$$a_h^* = \cA^*(s,r,h,\bfc_h,q_h^*) = \cA(k,\sigma(j),\bfc_h,q_h^*).$$
Next she calculates $\hat{a}_j$ to be
\begin{align*}
\hat{a}_j & = \sum_{h\in R_{\ell,j}} a_h = \sum_{h\in R_{\ell,j}} \cA(k,\sigma(j),\bfc_h,q_h^*)  \\
& = \cA(k,\sigma(j),\sum_{h\in R_{\ell,j}}\bfc_h,q_h^*) = \cA(k,\sigma(j),\bfx_\ell,q_h^*)  \\
& =  \cA(k,\sigma(j),\bfx_\ell,q_{\sigma(j)}). 
\end{align*}
Finally, by assigning
$$a_j' = \hat{a}_{\sigma^{-1}(j)} = \cA(k,\sigma(\sigma^{-1}(j)),\bfx_\ell,q_{\sigma(\sigma^{-1}(j))})= \cA(k,j,\bfx_\ell,q_j),$$
she completes with the last step of the algorithm. We can see that the privacy of the $s$ parts is kept since each server is required to invoke the algorithm $\cA$ with the parameter $\sigma(j)$. Since the permutation $\sigma$ was chosen uniformly at random, the distribution $\sigma(j)$ is identical for any choice of $\ell$, one of the $s$ parts of the database where Alice wants to retrieve a bit. Lastly, the privacy of the bit that Alice reads is guaranteed from the privacy of the PIR protocol $\cP$.

The coded PIR protocol $\cP^*$ uploads $D(\cP; n/s,k)$ bits to each server and downloads $D(\cP; n/s,k)$ from each server. Thus, we get that  $U^*(\cP;n,m,s) = m\cdot U(\cP;n/s,k)$ and $D^*(\cP;n,m,s) = m\cdot D(\cP;n/s,k)$.
\end{IEEEproof}

\section{Constructions of Coded PIR Schemes}\label{sec:cons}

In this section we give several methods to construct $k$-server PIR codes with the properties specified in Section~\ref{sec:coded_PIR}. As we shall see the properties of $k$-server PIR codes are similar to some of the existing codes in the literature, such as one-step majority logic codes~\cite{Costello}, codes with locality and availability~\cite{HYUS15, GHSY12, PHO13, RPDV14, TB14}, and combinatorial objects such as Steiner systems. We point out that the simple-parity code is the optimal 2-server PIR code and thus $A(s,2)=s+1$. Therefore, we focus in this section only on the case where $k>2$.

\subsection{The Cubic Construction}\label{subsec:cubic}
Our first construction is based on the geometry of multidimensional cubes.  Let us assume that $s=\sigma^{k-1}$ for some positive integer $\sigma$. We will give a construction of an $[m,s]$ $k$-server systematic PIR code, where $m=\sigma^{k-1} + (k-1) \sigma^{k-2} = s + (k-1) s^{(k-2)/(k-1)}$. This code will be denoted by $\cC_A(\sigma,k)$.

The information bits in $\cC_A$ will be denoted by $x_{i_1,i_2,\cdots,i_{k-1}}$, where $1\leq i_j\leq\sigma$ for $j\in [k-1]$. The $(k-1) \sigma^{k-2}$ redundancy bits, which are partitioned into $k-1$ groups, are denoted and defined as follows:
\begin{align*}
&p_{i_1,i_2,\cdots,i_{\xi-1},i_{\xi+1},\cdots,i_{k-1}}^{(\xi)}=\sum_{i_{\xi}=1}^{\sigma} x_{i_1,i_2,\cdots,i_{k-1}},
\end{align*}
for $\xi \in [k-1]$.

In the next example we demonstrate the construction of the code $\cC_A(\sigma,3)$.
\begin{example}\label{example:cubic construction}
Assume that $k=3$ and $s=\sigma^2$ for some positive integer $\sigma$. The code $\cC_A(\sigma,3)$ has in this case $2\sigma$ redundancy bits. The codewords are represented in a square array of size $(\sigma+1)\times (\sigma+1)$, without the bit in the bottom right corner. The information bits are stored in a $\sigma\times\sigma$ subsquare, and the remaining bits are the redundancy bits such that all rows and all columns are of even weight; see Fig.~\ref{fig:cubic}. So, for $i\in [\sigma]$,
\begin{align*}
&p_{i}^{(1)}=\sum_{j=1}^{\sigma} x_{i,j}\;, \\
&p_{i}^{(2)}=\sum_{j=1}^{\sigma} x_{j,i}.
\end{align*}
\begin{figure}[ht!]
\centering
\includegraphics[width=80mm]{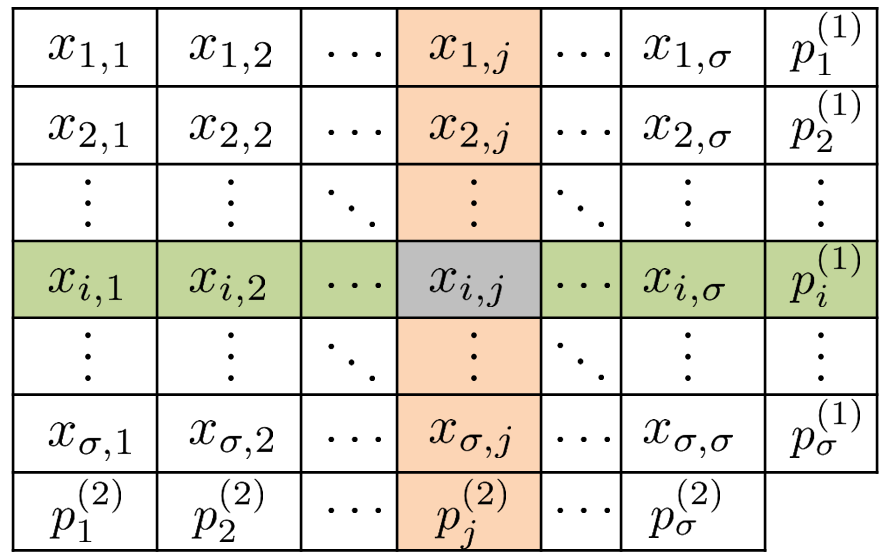}
\caption{The cubic construction for $3$-server PIR code. The bit $x_{i,j}$ can be recovered by itself, the bits in the $i$-th row besides $x_{i,j}$, and the bits in the $j$-th column besides $x_{i,j}$.}
\label{fig:cubic}
\end{figure}

One can verify that every information bit, $x_{i_1,i_2}$ for $i_1,i_2\in [\sigma]$, has three mutually disjoint sets such that $x_{i_1,i_2}$ is a linear function of the bits in each set. These sets are $\{x_{i_1,i_2}\}$, $\{x_{i_1,1},\ldots,x_{i_1,i_2-1},x_{i_1,i_2+1},\ldots,x_{i_1,\sigma}, p_{i_1}^{(1)}\}$, and $\{x_{1,i_2},\ldots,x_{i_1-1,i_2},x_{i_1+1,i_2},\ldots,x_{\sigma,i_2},p_{i_2}^{(2)}\}$.
Note that the cell in the bottom right corner was removed since it is not used in any of the recovering sets. Finally, we conclude from this example that \begin{align*}
\cA(\sigma^2,3)\leq  \sigma^2+2\sigma,
\end{align*}
and the storage overhead is $1+\frac{2}{\sigma}$, which approaches $1$ when $\sigma$ becomes large enough.
\end{example}

Next, we explicitly prove that the code $\cC_A(\sigma,k)$ is a $k$-server PIR code.
\begin{theorem}\label{thm:cubic bound}
For two positive integers $\sigma$ and $k$, the code $\cC_A(\sigma,k)$ is a $k$-server PIR code. In particular, we get that for any positive $s$
$$A(s,k)\leq s+(k-1)\ceil{s^{\frac{1}{k-1}}}^{k-2}.$$
\end{theorem}
\begin{IEEEproof}
For any information bit $x_{i_1,i_2,\cdots,i_{k-1}}$ in the code $\cC_A(\sigma,k)$, the following $k$ sets
\begin{align*}
&R_{i_1,\cdots,i_{k-1}}^{(0)}=\{x_{i_1,i_2,\cdots,i_{k-1}}\},\\
&R_{i_1,\cdots,i_{k-1}}^{(\xi)}=\{x_{i_1,\cdots,i_{\xi-1},\delta,i_{\xi+1},\cdots,i_{k-1}}:\delta \neq i_{\xi}\}\cup\{p_{i_1,\cdots,i_{\xi-1},i_{\xi+1},\cdots,i_{k-1}}^{(\xi)}\},\\
&\text{for } \xi=1,2,\cdots,k-1,
\end{align*}
are all disjoint, and for all $1\leq \xi \leq k-1$ we have
$$x_{i_1,i_2,\cdots,i_{k-1}}=\sum_{x\in R_{i_1,\cdots,i_{k-1}}^{(\xi)}} x.$$
Therefore, $\cC_A(\sigma,k)$ is a $k$-server PIR code.

For the general construction of arbitrary values of $s$, let $\sigma$ be such that $(\sigma-1)^{k-1}< s \leq \sigma^{k-1}$. Using the code $\cC_A(\sigma,k)$, we add $k-1$ sets, each of $\sigma^{k-2}$ redundancy bits, to the information bits to form a $k$-server PIR code. In case that $s<\sigma^{k-1}$, we simply treat the missing bits in the square as zeros. Hence,
\begin{align*}
A(s,k)\leq s+ (k-1)\sigma^{k-2}=s+(k-1)\ceil{s^{\frac{1}{k-1}}}^{k-2}.
\end{align*}
\end{IEEEproof}

For a fixed $k$, the asymptotic behavior of the storage overhead in the cubic construction is given by $1+\cO(s^{-\frac{1}{k-1}})$, which already proves that the asymptotic storage overhead approaches 1, that is,
$$\lim_{s\rightarrow \infty}\frac{A(s,k)}{s} = 1.$$
However, as we shall see in the sequel, it is still possible to improve the value of $A(s,k)$ for specific values of $s$ and $k$, and to find constructions which their storage overhead approaches 1 faster than the decay exponent given by the cubic construction.
Lastly, we note that a recursive form of this construction appears in~\cite{IKOS04} for the purpose of constructing batch codes.

\subsection{PIR Codes Based on Steiner Systems}\label{subsec:steiner_systems}
The idea behind a construction of any $k$-server PIR code $\cC$ is to form, for every information bit, $k$ mutually disjoint subsets of $[m]$, such that the information bit can be recovered by a linear combination of the bits in each set. Assume that $\cC$ is a systematic $[m,m-r=s]$ $k$-server PIR code. Then, we can partition its bits into two parts; the first one consists of the $s$ information bits, denoted by $x_1,\ldots, x_s$ and the second one is the $r$ redundancy bits $p_1,\ldots, p_r$, where every redundancy bit $p_i$ is characterized by a subset $S_i\subseteq [s]$ such that $p_i = \sum_{j\in S_i}x_j$.

According to this representation of systematic codes, every collection $\cS = (S_1,\ldots, S_r)$ of subsets of $[s]$ defines a systematic $[s+r,s]$ linear code $\cC_B(\cS)$. In the next lemma, we give sufficient (but not necessary) conditions such that the code $\cC_B(\cS)$ is a $k$-server PIR code.
\begin{lemma}\label{lem:systematic}
Let $\cS= (S_1,\ldots, S_r)$ be a collection of subsets of $[s]$, such that
\begin{enumerate}
\item For all $i\in [s]$, $i$ appears in at least $k-1$ subsets,
\item For all $j,\ell \in [r]$,  $|S_j \cap S_\ell|\leq 1$.
\end{enumerate}
Then, $\cC_B(\cS)$ is a $k$-server PIR code.
\end{lemma}
\begin{IEEEproof}
For any information bit $x_i$, $i\in [s]$, according to the first condition there exist some $k-1$ subsets $S_{i_1},\ldots,S_{i_{k-1}}$, such that $i\in S_{i_j}$ for $j\in [k-1]$. For each $j\in [k-1]$, let $R_j$ be the set $R_j = \{x_\ell \ : \ \ell \in S_{i_j}, \ell\neq i\} \cup \{p_{i_j}\}$, and finally let $R_k = \{x_i\}$. According to the second condition all these $k$ sets are mutually disjoint. Finally, it is straightforward to verify that $x_i$ is the sum of the bits in every set, and thus $\cC_B(\cS)$ is a $k$-server PIR code.
\end{IEEEproof}

After determining the conditions in which the code $\cC_B(\cS)$ is a $k$-server PIR code, we are left with the problem of finding such collections of subsets. Our approach to fulfill the conditions stated in Lemma~\ref{lem:systematic} is to search for existing combinatorial objects in the literature. One such an object is a \emph{Steiner system}. A Steiner system with parameters $t$, $\ell$, $n$, denoted by $S(t,\ell,n)$, is an $n$-element set $S$ together with a set of $\ell$-element subsets of $S$ (called blocks) with the property that each $t$-element subset of $S$ is contained in exactly one block. It is also commonly known that the number of subsets in a Steiner system $S(t,\ell,n)$ is $\binom{n}{t}/\binom{\ell}{t}$ and every element is contained in exactly $\binom{n-1}{t-1}/\binom{\ell-1}{t-1}$ subsets.

In order to satisfy the conditions in Lemma~\ref{lem:systematic}, we chose Steiner systems with $t=2$ so the intersection of every two subsets contains at most one element. Furthermore, in a Steiner system $S(2,\ell,n)$, the number of subsets is $\binom{n}{2}/\binom{\ell}{2} = n(n-1)/\ell(\ell-1)$ and every element is contained in $(n-1)/(\ell-1)$ subsets. Thus, we conclude with the following theorem.
\begin{theorem}\label{thm:steiner systems}
If a Steiner system $S(2,\frac{s-1}{k-1}+1,s)$ exists, then there exists an $[m=s+r,s]$ $k$-server PIR code where $r=\frac{s(k-1)^2}{s+k-2}$. Thus, under this assumption we have
\begin{align}\label{eq:Steiner column}
A\left(s,k\right)\leq s+\frac{s(k-1)^2}{s+k-2}.
\end{align}
Moreover, if a Steiner system $S(2,k-1,r)$ exists, then we have a $k$-server PIR code with parameters $[m,s]=[r+\frac{r(r-1)}{(k-1)(k-2)},\frac{r(r-1)}{(k-1)(k-2)}]$. Thus,
\begin{align}\label{eq:Steiner row}
A\left(\frac{r(r-1)}{(k-1)(k-2)},k\right)\leq r+\frac{r(r-1)}{(k-1)(k-2)}.
\end{align}
\end{theorem}
\begin{IEEEproof}
Let $\cS$ be a Steiner system $S(2,\frac{s-1}{k-1}+1,s)$, so the number of subsets in $\cS$ is $$r= \frac{s(s-1)}{(\frac{s-1}{k-1}+1)\frac{s-1}{k-1}} = \frac{s(k-1)^2}{s+k-2},$$ and  every element is contained in $$\frac{s-1}{(s-1)/(k-1)} = k-1$$ subsets. We also have that the intersection of every two subsets contains at most one element, so the conditions in Lemma~\ref{lem:systematic} hold and $\cC_B(\cS)$ is a $k$-server PIR code. To prove the bound given in~(\ref{eq:Steiner row}), let $\tau=\binom{r}{2}/{\binom{k-1}{2}}$ be the number of $(k-1)$-element subsets of $S(2,k-1,r)$, and denote them by $\cS_1,\cS_2,\cdots,\cS_{\tau}\subset [r]$. Let us construct the dual Steiner system $S'(2,\frac{r-1}{k-2},\tau)$ which consists of $r$ $(\frac{r-1}{k-2})$-element subsets of $[\tau]$ denoted by $\cS'_1,\cS'_2,\ldots,\cS'_r$, and has the property that $\cS'_i=\{a|a\in [\tau], i \in \cS_a\}$. We now use the first statement in~(\ref{eq:Steiner column}) to construct the code $\cC_B(S')$. It is clear that the redundancy of $\cC_B(S')$ is given by $r$, and the code length is given by $r+\tau=r+\frac{r(r-1)}{(k-1)(k-2)}$.
\end{IEEEproof}
\vspace{-1ex}
\begin{example}\label{example:Steiner systems}
A finite projective plane of order $q$, with the lines as blocks, is an $S(2,q+1,q^2+q+1)$ Steiner system. Since $q+1 = \frac{(q^2+q+1)-1}{(q+2)-1}+1$, we conclude that there exists an $[s+r,s]$ $(q+2)$-server PIR code, with $s=q^2+q+1$ information bits and
$$r= \frac{(q^2+q+1)(q+1)^2}{q^2+q+1 + q+2-2} = \frac{(q^2+q+1)(q+1)^2}{(q+1)^2} = q^2+q+1$$
redundancy bits. Note that the storage overhead of this code is 2. 
\end{example}

In order to evaluate the bound~\ref{eq:Steiner row} in Theorem~\ref{thm:steiner systems}, one is required to figure out the existence of $S(t,\ell,n)$ in general. Indeed, Wilson Theorem claims that for a fixed $\ell$, and sufficiently large $n$, a Steiner system $S(2,\ell,n)$ exists given that the following two conditions (also known as divisibility conditions) are satisfied (see~\cite{W72_1, W72_2, W72_3} for more details):
\begin{enumerate}
\item $\frac{n(n-1)}{\ell (\ell -1)} \in \mathbb{Z}$, and
\item $\frac{n-1}{\ell-1} \in \mathbb{Z}$.
\end{enumerate}
Wilson Theorem guarantees the existence of $S(2,k-1,r)$ for infinitely many values of $r$. Hence, for a fixed $k$, there are arbitrary large values for $r$ such that the bound in~(\ref{eq:Steiner row}) holds. Hence, the redundancy behaves asymptotically according to $A(s,k) - s = O(s^{1/2})$, which improves upon the cubic construction.

									\subsection{One-step Majority Logic Codes}\label{subsec:one_step_majority_logic}

\emph{One-step majority logic decoding} is a method to perform fast decoding by looking at disjoint parity check constraints that only intersect on a single bit (see~\cite{Costello} - Chapter 8.) These parity check constraints correspond to the codewords in the dual code, and hence, for a linear code $[n,k,d]$, the goal is to find, for each $i\in [n]$, a set of codewords in the dual code that intersect only on the $i$-th bit. These codewords are said to be \emph{orthogonal} on the $i$-th bit. The maximum number of such orthogonal vectors in the dual code (for every bit) is denoted by $J$, and if $J=d-1$, then the code is called \emph{completely orthogonalizable}.

In other words, if an $[n,k]$ code has $J$ orthogonal vectors on the $i$-th coordinate for some $i\in [n]$, then its dual code $\cC^{\perp}$ has $k-1=J$ codewords that are orthogonal on coordinate $i$. Assume that these codewords are given by
\begin{align}\label{eq:k-server_majority}
&c^{\perp}_{j}=x_i+x_{j_1}+x_{j_2}+\cdots+x_{j_{p_j}}\;\;\;\;\;\; &\forall 1\leq j\in  J,
\end{align}
where the sets $\{i\}$, and $\{j_1,j_2,\ldots,j_{p_i}\}$ for $j\in [J]$ are mutually disjoint. Such $[n,k,d]$ code with $J$ orthogonal vectors for each $i\in[n]$ is called a \emph{one-step majority logic code with $J$ orthogonal vectors}. Note that the definition of one-step majority logic codes is almost identical to the one of PIR codes given in Definition~\ref{def:PIR code}. While one-step majority logic codes guarantee that orthogonal vectors (or mutually disjoint sets) exist for all the bits in the code, in PIR codes we require this property only for the $s$ information bits. While it is not always straightforward to construct an appropriate generator matrix from a given code such that the $k$-server PIR property holds, for the case of one-step majority logic codes, we can always pick a systematic generator matrix and hence the PIR property follows. Lastly we note that the idea of using one-step majority logic codes was motivated by the recent work on codes for locality and availability in~\cite{HYUS15}.

We demonstrate the construction of such codes in the following example.
\begin{example}\label{ex:(15,7) cyclic code}
Consider a $(15,7)$ cyclic code generated by the polynomial
\begin{align*}
g(x)=1+x^4+x^6+x^7+x^8.
\end{align*}
The parity-check matrix of this code in the systematic form is given by
\[
H=
\left[\arraycolsep=4.2pt\def\arraystretch{1} {\begin{array}{cccccccccccccccc}

1 &0 &0 &0 &0 &0 &0 &0 &1 &1 &0 &1 &0 &0 &0 \\
0 &1 &0 &0 &0 &0 &0 &0 &0 &1 &1 &0 &1 &0 &0 \\
0 &0 &1 &0 &0 &0 &0 &0 &0 &0 &1 &1 &0 &1 &0 \\
0 &0 &0 &1 &0 &0 &0 &0 &0 &0 &0 &1 &1 &0 &1 \\
0 &0 &0 &0 &1 &0 &0 &0 &1 &1 &0 &1 &1 &1 &0 \\
0 &0 &0 &0 &0 &1 &0 &0 &0 &1 &1 &0 &1 &1 &1 \\
0 &0 &0 &0 &0 &0 &1 &0 &1 &1 &1 &0 &0 &1 &1 \\
0 &0 &0 &0 &0 &0 &0 &1 &1 &0 &1 &0 &0 &0 &1 \\
\end{array} } \right]
.\]
We observe that the following codewords in $C^{\perp}$
\begin{align*}
h_{3}=&(0 \;\;0\;\; 0 \;\;1 \;\;0 \;\;0 \;\;0 \;\;0 \;\;0 \;\;0 \;\;0 \;\;1 \;\;1 \;\;0 \;\;1),\\
h_{1+5}=&(0\;\; 1\;\; 0 \;\;0 \;\;0 \;\;1 \;\;0 \;\;0 \;\;0 \;\;0 \;\;0 \;\;0 \;\;0 \;\;1 \;\;1),\\
h_{0+2+6}=&(1\;\; 0\;\; 1\;\; 0\;\; 0\;\; 0\;\; 1\;\; 0\;\; 0\;\; 0\;\; 0\;\; 0\;\; 0\;\; 0\;\; 1),\\
h_{7}=&(0\;\; 0\;\; 0\;\; 0\;\; 0\;\; 0\;\; 0\;\; 1\;\; 1\;\; 0\;\; 1\;\; 0\;\; 0\;\; 0\;\; 1),
\end{align*}
are orthgonal on coordinate $14$. That gives us five mutually disjoint sets $\{3,11,12\}$, $\{1,5,13\}$, $\{0,2,6\}$, $\{7,8,10\}$, and $\{14\}$ that are required in  Definition~\ref{def:PIR code} to make five different queries on server $14$. The same statement is correct for all other coordinates due to the cyclicity of the code. So, $\cC$ is a $5$-server PIR code. The storage overhead of the coded PIR scheme based on $\cC$ is given by $\frac{1}{\text{code rate}}=\frac{15}{7}$, which is significantly better than the uncoded PIR.
\end{example}

There are several algebraic constructions for one-step majority logic codes. However, the explicit relation between the code length and redundancy is only known for a few of them. \emph{Type-$1$ Doubly Transitive Invariant (DTI) Codes} (see~\cite{Costello} - p. 289) are cyclic codes with \emph{almost} completely orthogonalizable property. An explicit relation between the code length $m=(2^M-1)$, code dimension $(s)$, and the number of orthogonal codewords in the dual code $(J)$, is known for specific choices of these parameters:
\begin{itemize}
\item
Case I. Let $\theta,\ell$ be two positive integers. For $M=2\theta\ell$ and $J=2^{\ell}+1$, the redundancy of the type-$1$ DTI code of length $m$ is given by
\begin{align}\label{eq:DTI case 1}
r=(2^{\theta+1}-1)^{\ell}-1.
\end{align}
We refer to these codes by $\cC_{C_1}(\theta,\ell)$.
\item Case II. Let $\lambda,\ell$ be two positive integers. For $M=\lambda\ell$ and $J=2^{\ell}-1$, the redundancy of the type-$1$ DTI code of length $m$ is given by
\begin{align}\label{eq:DTI case 2}
r= 2^M - (2^{\lambda}-1)^{\ell}-1.
\end{align}
We refer to these codes by $\cC_{C_2}(\lambda,\ell)$.
\end{itemize}
We refer the reader to ~\cite{Costello} for the algebraic construction and the calculation method used in deriving these parameters.

\begin{theorem}\label{thm:majority logic}
For any positive integers $\theta, \ell$, and $\lambda$, the Type-$1$ DTI codes $\cC_{C_1}(\theta,\ell)$, $\cC_{C_2}(\lambda,\ell)$ are $(2^{\ell}+2)$-server, and $2^{\ell}$-server PIR codes, respectively. In particular, we get that
\begin{align*}
&A(2^{2 \theta \ell}-(2^{\theta+1}-1)^{\ell},2^{\ell}+2)\leq 2^{2\theta \ell} -1,\\
&A((2^{\lambda}-1)^{\ell}-1,2^{\ell})\leq 2^{\lambda \ell} -1,
\end{align*}
and hence for any fixed $k$, there exists a family of $k$-server PIR codes with asymptotic storage overhead of $1+\cO(s^{-\frac{1}{2}})$.
\end{theorem}

\begin{IEEEproof} 
We have already shown that a one-step majority logic code with $J$ orthogonal vectors, is also a $(J+1)$-server PIR code. So we are left with only calculating the code dimensions according to the redundancies in~(\ref{eq:DTI case 1}) and~(\ref{eq:DTI case 2}).
\begin{itemize}
\item For $\cC_{C_1}(\theta,\ell)$, the code dimension is given by $s=m - r = 2^{2 \theta \ell}-(2^{\theta+1}-1)^{\ell}$.
\item For $\cC_{C_2}(\lambda,\ell)$, the code dimension is given by $s=m - r = (2^{\lambda}-1)^{\ell}-1$.
\end{itemize}
So the upper bounds are validated. For the asymptotic analysis, we point out that for a given fixed $J$, as the number of servers grows, the rates of the codes in both cases I, and II become arbitrary close to $1$. In particular, when $k$ is fixed and $s$ becomes large, the storage overhead in $\cC_{C_2}(\lambda,\ell)$ is
\begin{align*}
\frac{2^{\lambda \ell}-1}{(2^{\lambda}-1)^{\ell}-1}  \approx (\frac{2^{\lambda}}{2^{\lambda}-1})^{\ell} \approx 1 + \frac{\ell}{2^{\lambda}-1} = 1+ O(s^{-\frac{1}{\ell}}),
\end{align*}
which is  an improvement compared to Theorem~\ref{thm:cubic bound} in the asymptotic regime. An even better storage overhead is achieved by $\cC_{C_1}(\theta,\ell)$ codes in the asymptotic regime:
\begin{align}\label{eq:case I - asymptotic}
\frac{2^{2\theta \ell}-1}{2^{2 \theta \ell }- (2^{\theta+1}-1)^{\ell}} = 1 + O (s^{-\frac{1}{2}}).
\end{align}
\end{IEEEproof}

Note that this construction not only outperforms the former ones with respect to the upper bound on the asymptotic storage overhead, but also gives a bound on $A(s,k)$ that does not depend on $k$ is in the asymptotic regime. Considering that the construction based on Steiner systems also result in a similar bound, we ask the following two questions regarding the asymptotic storage overhead behavior of $k$-server PIR codes.\\
{\vspace{1ex} \bf Question 1.} Is~(\ref{eq:case I - asymptotic}) the optimal asymptotical behavior for $A(s,k)$? \\
A more challenging question would be to show the same statement for finite numbers. In particular, \\
{\vspace{0ex} \bf Question 2.} Are there any values of $s$ and $k\geq 3$ such that $A(s,k) < s+ \sqrt{s}$?

\subsection{Constant-Weight Codes}\label{subsec:constant_weight}

Assume that $G$ is a generator matrix of a systematic $k$-server PIR code $\cC$  of length $m$ and dimension $s$. We rewrite $G$ as
\begin{align}\label{eq:systematic_construction}
G=[I_{s} | M_{s\times r}],
\end{align}
where $I_s$ is the $s\times s$ identity matrix and $M_{s\times r}$ corresponds to the $r$ parities in $\cC$. Let us look at the systematic PIR codes from a graph theory point of view by interpreting $M$ as the incidence matrix of a bipartite graph $\cG$ with partite sets $\cX=\{x_1,x_2,\ldots,x_s\}$ and $\cP=\{p_1,p_2,\ldots,p_r\}$, and edges $\cE=\{\{x_i,p_j\}|M_{ij}=1\}$. We call $\cC$ by the \emph{Systematic PIR code based on $\cG$.} The following lemma is an equivalent statement to Lemma~\ref{lem:systematic}.

\begin{lemma}\label{lemma:cycle_free_construction}
Let $\cG$ be a bipartite graph with partite sets $\cX=\{x_1,x_2,\ldots,x_s\}$, $\cP=\{p_1,p_2,\ldots,p_r\}$, and the incidence matrix $M$, where $k-1=\min_{x\in \cX} \deg(x)$. Further, assume that $\cG$ has no cycles of length $4$. If $\cC$ is the systematic code based on $\cG$ with generator matrix defined in in~(\ref{eq:systematic_construction}), then $\cC$ is a $k$-server PIR code of length $m=s+r$ and dimension $s$.
\end{lemma}
\begin{IEEEproof}
Consider $x_i$ and $k-1$ of its parity neighbors $\{p_{i_1},p_{i_2},\ldots,p_{i_{k-1}}\}\subset \cP$. Let $R_{i}^j\subset \cX$ denote the neighbor set of $p_{i_j}$. Since $\cG$ is $4$-cycle free, the sets $R_i^j \setminus \{x_i\}$ (for a fixed $i$ and $j\in[k-1]$) are mutually disjoint. It is also easy to see that $\{x_i\}$, and $\{p_{i_j}\}\cup R_i^j \setminus \{x_i\}$ (for $j=1,2,\ldots,k-1$) form $k$ disjoint recovery sets for $u_i$. In other words,
\begin{align*}
p_{i_j}=\sum_{x_{\alpha}\in R_i^j} x_{\alpha} \Rightarrow x_i=p_{i_j}+ \sum_{x_{\alpha}\in R_i^j \setminus \{x_i\}} x_{\alpha} &\;\;\;\;\;\;\; \text{ for } j=1,2,\cdots k-1.
\end{align*}
\end{IEEEproof}

Now we are ready to proceed to the final construction of $k$-server PIR codes, which will be first demonstrated by an example.
\begin{example}\label{ex:constant weight codes}
Consider the $3$-server PIR code $\cC$ given by the systematic generator matrix
\
 \[
   G= [I_{10} | M_{10\times 5}] =
  \left[ {\begin{array}{ccccccccccccccc}
   1 & 0 & 0 & 0 & 0 & 0 & 0 & 0 & 0 & 0 & 1 & 1 & 0 & 0 & 0 \\
   0 & 1 & 0 & 0 & 0 & 0 & 0 & 0 & 0 & 0 & 1 & 0 & 1 & 0 & 0 \\
   0 & 0 & 1 & 0 & 0 & 0 & 0 & 0 & 0 & 0 & 1 & 0 & 0 & 1 & 0 \\
   0 & 0 & 0 & 1 & 0 & 0 & 0 & 0 & 0 & 0 & 1 & 0 & 0 & 0 & 1 \\
   0 & 0 & 0 & 0 & 1 & 0 & 0 & 0 & 0 & 0 & 0 & 1 & 1 & 0 & 0 \\
   0 & 0 & 0 & 0 & 0 & 1 & 0 & 0 & 0 & 0 & 0 & 1 & 0 & 1 & 0 \\
   0 & 0 & 0 & 0 & 0 & 0 & 1 & 0 & 0 & 0 & 0 & 1 & 0 & 0 & 1 \\
   0 & 0 & 0 & 0 & 0 & 0 & 0 & 1 & 0 & 0 & 0 & 0 & 1 & 1 & 0 \\
   0 & 0 & 0 & 0 & 0 & 0 & 0 & 0 & 1 & 0 & 0 & 0 & 1 & 0 & 1 \\
   0 & 0 & 0 & 0 & 0 & 0 & 0 & 0 & 0 & 1 & 0 & 0 & 0 & 1 & 1 \\
  \end{array} } \right].
\]
The corresponding bipartite graph is also shown in Fig.~\ref{pic:bipartite graph}.

\begin{figure}[h]
\centering
\includegraphics[width=90mm]{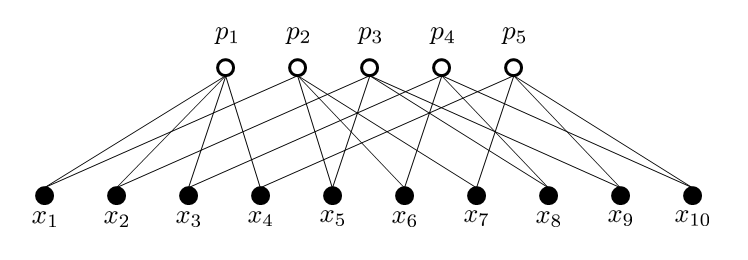}
\caption{The bipartite graph $\cG$ associated with the matrix $M_{10\times 5}$.  \label{pic:bipartite graph}}
\end{figure}
We observe that $\deg(x_i)=2$ for all $i$ as expected since we only need $k-1=2$ elements in $\cN(x_i)$ to recover $x_i$, where $\cN(\alpha)$ is the neighborhood set of $\alpha$. Moreover, 
\begin{align}\label{eq:constant_weight_condition}
|\cN(x_i)\cap \cN(x_j)|\leq 1\text{ for }i \neq j,
\end{align} 
which guarantees that the recovering sets for $x_i$ are mutually disjoint.  
\end{example}

The requirement that $\deg(x_i)\geq k-1$ can be replaced with $\deg(x_i)=k-1$ in this construction. This motivates us to look at constant-weight codes where the codewords are all rows in the matrix $M$, such that the condition in~(\ref{eq:constant_weight_condition}) still holds. For instance, we look at a constant weight code with weight $k-1$ and minimum distance $2k-4$, which guarantees the condition in~(\ref{eq:constant_weight_condition}). Let $M(k,r)$ be the list of the largest code  of length $r$ whose codewords have weight $k-1$ and their minimum distance is $2k-4$. $\cC_D(k,r)$ is a $k$-server PIR code defined by its systematic generator matrix $G_{(k,r)}=[I|M(k,r)]$.

We use the notation $B(n,w,d)$ to denote the maximum number of codewords of length $n$ and weight $w$ with minimum distance $d$. There are numerous works and studies aiming to determine the precise values of $B(n,w,d)$ in general, but the explicit formula is only found for the trivial cases. A complete collection of the known precise values and both upper bounds and lower bounds on $B(n,w,d)$ is given in~\cite{AEB}.

\begin{theorem}\label{thm:constant weight codes}
For any $k$ the code $\cC_D(k,r)$ is a $k$-server PIR code. In particular, we get that for any positive integer $k$
\begin{align*}
A\big (B(r,k-1,2k-4),k\big ) \leq B(r,k-1,2k-4)+r.
\end{align*}
\end{theorem}
\begin{IEEEproof}
Let $\cG$ be the bipartite graph whose incidence matrix is $M(k,r)$. It is clear that $\deg(x)=k-1$ for all $x\in \cX$. Also, 
\begin{align*}
&|\cN(x)| = |\cN(y)| = k-1, \\
&|\{\cN(x) \cup \cN(y)\}\setminus \{\cN(x) \cap \cN(y)\}|\geq 2k-4\\
&\Rightarrow \{\cN(x) \cap \cN(y)\}| \leq 1.
\end{align*}
Hence, all of the conditions in Lemma~\ref{lemma:cycle_free_construction} are satisfied and $\cC_D(k,r)$ is a $k$-server PIR code. To validate the parameters in the theorem it suffices to note that $|M(k,r)|=B(r,k-1,2k-4)$, so we have $B(r,k-1,2k-4)$ rows in $G$, and $r$, the length of the codewords in $M(k,r)$, determines the redundancy.
\end{IEEEproof}
\begin{example}\label{ex:constant weight codes of weight 2}
The only known explicit formula for $B(n,w,d)$ is when $w=2$ and $d=2$. It is easy to see that any two different codewords of weight $2$ have distance at least $2$ as well. Hence $B(n,2,2)= \binom{n}{2}$. So,
\begin{align}\label{eq:constant weight codes of weight 2}
A\left( \binom{n}{2}, 3 \right) \leq n + \binom{n}{2}.
\end{align}
\end{example}

According to inequality~\ref{eq:constant weight codes of weight 2}, we observe again that the asymptotic behavior of $A(s,3) - s$ is $O(s^{1/2})$.  
The construction based on Steiner systems and the last construction based upon constant-weight codes are both equivalent to the problem of finding bipartite graphs with $s$ vertices on the left and $r$ vertices on the right, where all the left vertices have degree $k$, the graph has girth at least 6, while minimizing the value of $r$. Clearly, if a Steiner system with the desired parameter exists, then it is an optimal solution. However, constant-weight codes provide a solution in the general case particularly when the desired Steirner system does not exist. The following theorem on bipartite graphs is both well-known and trivial (see for example Proposition 7 in~\cite{N08}), and shows that by using this method of construction for $k$-server PIR codes, we can not achieve a better asymptotic storage overhead than the one achieved by Steiner systems. We include here the proof for the sake of completeness of the results in the paper. 

\begin{theorem}\label{thm:sqrt of s}
Let $\cG$ be a bipartite graph with partite sets $\cX=\{x_1,x_2,\ldots,x_s\}$ and $\cP=\{p_1,p_2,\ldots,p_r\}$, where $\deg(x)=k>2$ for all $x\in \cX$. If the graph has girth at least 6, then $r(r-1) \geq sk(k-1)$. Hence, for a fixed $k$, we have $r = O(s^{1/2})$.
\end{theorem}
\begin{IEEEproof}
Let $\cN(x)$ denote the neighbor set of the vertex $x$. Since the graph has no $4$-cycles, then there are no $i,j, \in [s]$ and $a,b\in [r]$, such that
\begin{align*}
&\{p_a,p_b\} \subset \cN(x_i),\text{ and }\\
&\{p_a,p_b\} \subset \cN(x_j).
\end{align*}
Therefore, 
\begin{align*}
\binom{r}{2} \geq \sum_{i\in [s]} \binom{|\cN(x_i)|}{2} = s \binom{k}{2}.
\end{align*}
\end{IEEEproof}

\section{Optimal Storage Overhead for fixed $s$ and $k$}\label{sec:small_s_k}

In this section, we study the value of $A(s,k)$ for small $s$ and $k$, in particular we give an upper bound on $A(s,k)$ for all values of $s\leq 32$ and $k\leq 16$. In order to give the best upper bounds, we benefit from a few supplementary lemmas that together with the constructions introduced in section~\ref{sec:cons} form a recursive method in deriving the upper bounds on $A(s,k)$. 

Note that the constructions introduced in Section~\ref{sec:cons} do not cover all values of $s$ and $k$.  The following lemmas give simple tools to derive upper bounds for all values of $s$ and $k$.

\begin{lemma}\label{lem:sublinearity}
We have the following inequalities for all non-negative integer values of $s$, $k$, $s'$, and $k'$:
\begin{enumerate}[(a)]
\item $ A(s,k+k') \leq A(s,k)+A(s,k')$,
\item $ A(s+s',k) \leq A(s,k)+A(s',k)$,
\item $ A(s,k) \leq A(s,k+1)-1$,
\item $ A(s,k) \leq A(s+1,k)-1$.
\end{enumerate}
\end{lemma}

\begin{table*}[t]
\caption{Upper bound for $A(s,k)$ for small values of $s$ and $k$. For each $k$, the value on the left represents the size of the best PIR code constructions i.e. $A(s,k)$, and the right column represents the storage overhead $\frac{A(s,k)}{s}$ associated with that construction. By lemma~\ref{lem:even odd} the value of $A(s,k)$ for odd $k$ is given by $A(s,k+1)-1$. Starred values are proved to be optimal. }
\label{table:small values}
\resizebox{1\textwidth}{!}{
\centering
\begin{tabular}{ c || c |c|c|c|c|c|c| c|c |c|c|c|c|c|c| c| c | c | c | c | c | c}
$\bf {s \backslash k}$	&\multicolumn{2}{c}{\bf 2}\vline	&\multicolumn{2}{c}{\bf 3}\vline	&\multicolumn{2}{c}{\bf 4}\vline	&\multicolumn{2}{c}{\bf 6}\vline &\multicolumn{2}{c}{\bf 8}\vline	&\multicolumn{2}{c}{\bf 10}\vline	&\multicolumn{2}{c}{\bf 12}\vline	&\multicolumn{2}{c}{\bf 14}\vline	&\multicolumn{2}{c}{\bf 16}\vline\\
\hline
\bf 1	&$2^*$	&2.00	&$3^*$	&3.00	&$4^*$	&4.00	&$6^*$	&6.00	&$8^*$	&8.00	&$10^*$	&10.0	&$12^*$	&12.0	&$14^*$	&14.0	&$16^*$	&16.0\\
\hline
\bf 2	&$3^*$	&1.50	&$5^*$	&2.50	&$6^*$	&3.00	&$9^*$	&4.50	&$12^*$	&6.00	&$15^*$	&7.50	&$18^*$	&9.00	&$21^*$	&10.5	&$24^*$	&12.0\\
\hline
\bf 3	&$4^*$	&1.33	&$6^*$	&2.00	&$7^*$	&2.33	&$11^*$	&3.67	&$14^*$	&4.67	&$18^*$	&6.00	&$21^*$	&7.00	&$25^*$	&8.33	&$28^*$	&9.33\\
\hline
\bf 4	&$5^*$	&1.25	&8	&2.00	&9	&2.25	&$12^*$	&3.00	&$15^*$	&3.75	&20	&5.00	&24	&6.00	&$27^*$	&6.75	&$30^*$&7.50\\
\hline
\bf 5	&$6^*$	&1.20	&10	&2.00	&11	&2.20	&13	&2.60	&19	&3.80	&24	&4.80	&26	&5.20	&29	&5.80	&$31^*$ &6.20\\
\hline
\bf 6	&$7^*$	&1.17	&11	&1.83	&12	&2.00	&14	&2.33	&21	&3.50	&26	&4.33	&28	&4.67	&35	&5.83	&40	&6.67\\
\hline
\bf 7	&$8^*$	&1.14	&12	&1.71	&13	&1.86	&15	&2.14	&23	&3.29	&28	&4.00	&30	&4.29	&38	&5.43	&43	&6.14\\
\hline
\bf 8	&$9^*$	&1.13	&13	&1.63	&14	&1.75	&20	&2.50	&28	&3.50	&34	&4.25	&40	&5.00	&48	&6.00	&54	&6.75\\
\hline
\bf 9	&$10^*$	&1.11	&14	&1.56	&15	&1.67	&23	&2.56	&30	&3.33	&38	&4.22	&45	&5.00	&53	&5.89	&60	&6.67\\
\hline
\bf 10	&$11^*$	&1.10	&17	&1.70	&18	&1.80	&24	&2.40	&35	&3.50	&41	&4.10	&48	&4.80	&57	&5.70	&61	&6.10\\
\hline
\bf 11	&$12^*$	&1.09	&19	&1.73	&20	&1.82	&25	&2.27	&37	&3.36	&42	&3.82	&50	&4.55	&62	&5.64	&67	&6.09\\
\hline
\bf 12	&$13^*$	&1.08	&20	&1.67	&21	&1.75	&26	&2.17	&39	&3.25	&43	&3.58	&52	&4.33	&64	&5.33	&69	&5.75\\
\hline
\bf 13	&$14^*$	&1.08	&21	&1.62	&22	&1.69	&27	&2.08	&41	&3.15	&44	&3.38	&54	&4.15	&66	&5.08	&71	&5.46\\
\hline
\bf 14	&$15^*$	&1.07	&22	&1.57	&23	&1.64	&29	&2.07	&43	&3.07	&45	&3.21	&58	&4.14	&68	&4.86	&74	&5.29\\
\hline
\bf 15	&$16^*$	&1.07	&23	&1.53	&24	&1.60	&34	&2.27	&44	&2.93	&46	&3.07	&62	&4.13	&70	&4.67	&80	&5.33\\
\hline
\bf 16	&$17^*$	&1.06	&24	&1.50	&25	&1.56	&37	&2.31	&45	&2.81	&47	&2.94	&64	&4.00	&72	&4.50	&84	&5.25\\
\hline
\bf 17	&$18^*$	&1.06	&27	&1.59	&28	&1.65	&38	&2.24	&46	&2.71	&48	&2.82	&66	&3.88	&76	&4.47	&86	&5.06\\
\hline
\bf 18	&$19^*$	&1.06	&28	&1.56	&29	&1.61	&39	&2.17	&47	&2.61	&49	&2.72	&68	&3.78	&78	&4.33	&88	&4.89\\
\hline
\bf 19	&$20^*$	&1.05	&29	&1.53	&30	&1.58	&40	&2.11	&48	&2.53	&50	&2.63	&70	&3.68	&80	&4.21	&90	&4.74\\
\hline
\bf 20	&$21^*$	&1.05	&30	&1.50	&31	&1.55	&41	&2.05	&49	&2.45	&51	&2.55	&72	&3.60	&82	&4.10	&92	&4.60\\
\hline
\bf 21	&$22^*$	&1.05	&31	&1.48	&32	&1.52	&42	&2.00	&50	&2.38	&52	&2.48	&74	&3.52	&84	&4.00	&94	&4.48\\
\hline
\bf 22	&$23^*$	&1.05	&32	&1.45	&33	&1.50	&47	&2.14	&51	&2.32	&53	&2.41	&76	&3.45	&86	&3.91	&100	&4.55\\
\hline
\bf 23	&$24^*$	&1.04	&33	&1.43	&34	&1.48	&50	&2.17	&52	&2.26	&54	&2.35	&78	&3.39	&88	&3.83	&104	&4.52\\
\hline
\bf 24	&$25^*$	&1.04	&34	&1.42	&35	&1.46	&51	&2.13	&53	&2.21	&55	&2.29	&80	&3.33	&90	&3.75	&106	&4.42\\
\hline
\bf 25	&$26^*$	&1.04	&35	&1.40	&36	&1.44	&52	&2.08	&54	&2.16	&56	&2.24	&82	&3.28	&92	&3.68	&108	&4.32\\
\hline
\bf 26	&$27^*$	&1.04	&38	&1.46	&39	&1.50	&53	&2.04	&55	&2.12	&57	&2.19	&84	&3.23	&96	&3.69	&110	&4.23\\
\hline
\bf 27	&$28^*$	&1.04	&39	&1.44	&40	&1.48	&54	&2.00	&56	&2.07	&58	&2.15	&86	&3.19	&98	&3.63	&112	&4.15\\
\hline
\bf 28	&$29^*$	&1.04	&40	&1.43	&41	&1.46	&55	&1.96	&57	&2.04	&59	&2.11	&88	&3.14	&100	&3.57	&114	&4.07\\
\hline
\bf 29	&$30^*$	&1.03	&41	&1.41	&42	&1.45	&56	&1.93	&58	&2.00	&60	&2.07	&90	&3.10	&102	&3.52	&116	&4.00\\
\hline
\bf 30	&$31^*$	&1.03	&42	&1.40	&43	&1.43	&57	&1.90	&59	&1.97	&61	&2.03	&92	&3.07	&104	&3.47	&118	&3.93\\
\hline
\bf 31	&$32^*$	&1.03	&43	&1.39	&44	&1.42	&58	&1.87	&60	&1.94	&62	&2.00	&94	&3.03	&106	&3.42	&120	&3.87\\
\hline
\bf 32	&$33^*$	&1.03	&44	&1.38	&45	&1.41	&59	&1.84	&61	&1.91	&63	&1.97	&96	&3.00	&108	&3.38	&122	&3.81\\
\hline

\end{tabular}
}
\end{table*}

\begin{IEEEproof}
To prove the inequality in \emph{(a)}, assume that $\cC$ and $\cC'$ are $k$-server and $k'$-server PIR codes with parameters $[m,s]$ and $[m',s]$, and their generator matrices are given by $G$ and $G'$, respectively. It is easy to see that the concatenation of $\cC$ and $\cC'$ is a $(k+k')$-server PIR code with parameters $[m+m',s]$ and its generator matrix is given by $G_{\text{conc}}=[G \; | \; G']$. To prove the inequality in \emph{(b)}, assume again that $\cC$ and $\cC'$ are $k$-server PIR codes with parameters $[m,s]$ and $[m',s']$, and their generator matrices are given by $G$ and $G'$, respectively. The direct sum code (also known as the product code) of $\cC$ and $\cC'$ is a $k$-server PIR code with parameters $[m+m',s+s']$ whose generator matrix is given by
\[
   G^*=
  \left[ {\begin{array}{cc}
   G & 0_{s\times m'} \\
   0_{s' \times m} & G' \\
  \end{array} } \right].
\]
To prove \emph{(c)}, let us assume that $\cC$ is a $(k+1)$-server PIR code with parameters $[m,s]$ and a generator matrix $G$. According to Definition~\ref{def:PIR code}, for every information bit $u_i, i\in [s]$, there exist $k+1$ mutually disjoint sets $R_{i,1},\cdots,R_{i,k+1}\subset [m]$ such that for all $j\in[k]$, $u_i$ is a linear function the bits in $R_{i,j}$. It is now clear that deleting one of the coordinates from $G$ or equivalently puncturing the code $\cC$ in one of its coordinates can truncate at most one of these disjoint recovery sets. Hence the punctured code $\cC_{\text{punc}}$ whose parameters are given by $[m-1,s]$ is a $k$-server PIR code. We postpone the proof of part \emph{(d)} to the end of this section, where we discuss whether Definition~\ref{def:PIR code} is a property of the generator matrix or it can be interpreted as a property of the code itself.
\end{IEEEproof}

\begin{lemma}\label{lem:even odd}
If $k$ is odd, then $A(s,k+1)=A(s,k)+1$.
\end{lemma}
\begin{IEEEproof}
Utilizing part (c) in Lemma~\ref{lem:sublinearity}, it only suffices to show that if $k$ is odd, then $A(s,k+1) \leq A(s,k)+1$. To do so, assume that $\cC$ is a $k$-server PIR code with parameters $[m,s]$ and generator matrix $G$. For any $i\in[s]$ we should be able to find $k$ disjoint subsets of columns where the columns in each subset sum up to the vector $\bfe_i$. If the sum of all columns in $G$ is $\zero$ ,then clearly the sum of the remaining columns (the ones that are left out of the $k$ subsets) is also the vector $\bfe_i$. Hence the code is actually a $(k+1)$-server PIR code and we are done. If not, append one more column to $G$ so that the sum of all the columns is $\zero$. Then the resulting matrix is a generator matrix for a $(k+1)$-server PIR code. 
\end{IEEEproof}

By selecting the best constructions for $A(s,k)$ from section~\ref{sec:cons} for each individual $s$ and $k$, and then updating the table with respect to lemmas~\ref{lem:sublinearity} and~\ref{lem:even odd}, we are finally able to give an upper bound on $A(s,k)$ for all values of $s$, and $k$. Table~\ref{table:small values} contains the upper bound obtained on $A(s,k)$ for all $k \leq 16$ and $s\leq 32$. We observe that the storage overhead is significantly improved compared to the traditional uncoded PIR scheme. Moreover, the inequality $A(s,k)\geq s+\sqrt{s}$ always hold. The asymptotic behavior of $A(s,k)$ is discussed in next section.

In the remainder of this section, we seek to address the following key question: Is it the generator matrix that has the $k$-server PIR property, or it can be interpreted as a property of the code? Let us begin the discussion with the following definition.

\begin{definition}\label{def:B_k}
We say that an $[m,m-s]$ binary linear code $C$ has property $B_k$ if there exist $s$ cosets of $C$ such that:
\begin{enumerate}[a)]
\item Every coset contains $k$ disjoint vectors, and
\item The linear span of these cosets is the entire space $\F_2^m$.
\end{enumerate}
\end{definition}

\begin{theorem}\label{thm:B_k property}
If the code $C$ has the property $B_k$, then its dual code is a $k$-server PIR code.
\end{theorem}
\begin{IEEEproof}
We show that the definition~\ref{def:B_k} and~\ref{def:PIR code} are equivalent. Clearly, Definition~\ref{def:B_k} above is a property of a code, not a matrix. Now, given a generator matrix $G$ with for the PIR code $C'$, we get a code $C$ with property $B_k$ by simply taking $C$ to be the code defined by $G$ as its parity-check matrix. Now lets proceed to the other direction.

Assume that a code $C$ with property $B_k$ is given. Let $C_1,C_2,\cdots,C_s$ be the $s$ linearly independent cosets of $C$, each containing $k$ disjoint vectors. Start with an arbitrary parity-check matrix $H$ for $C$. Let $\sigma_1,\sigma_2,\cdots,\sigma_s$ denote the syndromes of $C_1,C_2,\dots,C_s$ with respect to $H$. Let $S$ be the $s \times s$ matrix having these syndromes as its columns. Note that condition b) of Definition~\ref{def:B_k} guarantees that $S$ is full-rank. Now form the $s \times (m+s)$ matrix $[H|S]$, and perform elementary row operations on this matrix to get $[H'|S']$ where $S'$ is the $s \times s$ identity matrix. Then the matrix $H'$ is a generator matrix for the $k$-server PIR code $C'$, which is clearly the dual code of $C$.
\end{IEEEproof}

The following lemma from the theory of the linear codes is essential for the proof of part \emph{(d)} in Lemma~\ref{lem:sublinearity}. We leave the proof to the reader.

\begin{lemma}\label{lem:independent syndromes}
Let $C$ be an $(m,k)$ binary linear code. Given a positive $t \le m-k$, let $C_1,C_2,\dots,C_t$ be cosets of $C$, and let $s_1,s_2,\dots,s_t$ be their syndromes. Then
\begin{align*}
\dim \big(\text{span}(C_1,C_2,\dots,C_t)\big) = k + t
\end{align*}
if and only if the syndromes $s_1,s_2,\dots,s_t$ are linearly independent.
\end{lemma}

We are now able to prove part \emph{(d)} of Lemma~\ref{lem:sublinearity} as promised earlier.
\begin{IEEEproof}[Proof of Lemma~\ref{lem:sublinearity}.(d)]
Suppose $A(s,k)=m$. Then there exists an $[m,m-s]$ code $C$ with property $B_k$. Moreover, no column in a parity check matrix for $C$ is entirely zero, otherwise $A(s,k)\leq m-1$. Puncture the code $C$ in any position. Upon puncturing, a) above remains true trivially. It remains to show that we can find some $s-1$ cosets of the punctured code that generate $\F_2^{m-1}$, which is a direct result of Lemma~\ref{lem:independent syndromes}. Hence the resulting $[m-1,m-s]$ code has property $B_k$, and it is a $k$-server PIR code.
\end{IEEEproof}

\section{Asymptotic Behavior of Coded PIR}\label{sec:asymp_behavior}
While deriving the precise values of $A(s,k)$ was our initial interest, studying the asymptotic behavior of $A(s,k)$ is no less interesting. In particular, lower bounds will help us to find constructions with the optimal storage overhead. Asymptotically, we will analyze the value of $A(s,k)$ when $s$ is fixed and $k$ is large, and vice versa. We briefly mention that we solved the first case while the lower bounds for the latter are yet to be found.

\subsection{Storage overhead for fixed $s$}

Let us first focus on the case where $s$, the ratio between the length of the whole data and the storage size of each server, is a fixed integer number, but $k$, the PIR protocol parameter, is large.

\begin{theorem}\label{thm:small_s}
For any pair of integer numbers $s$, and $k$, we have
\begin{align}\label{eq:lower_bound_small_s}
A(s,k)\geq \frac{2^s-1}{2^{s-1}}k,
\end{align}
with equality if and only if $k$ is divisble by $2^{s-1}$.
\end{theorem}

Let us use the following example to illustrate the proof.

\begin{example}\label{exm:lower_bound_s_3}
Assume $s=3$, and $C$ is an $(m,3)$ PIR code with $k$-server PIR property. The generator matrix of $C$ contains $m$ columns, each of length $3$. The list of all possible options is shown in table~\ref{table:example_s_3}. Let us assume that the column multiplicities are given by $\mu_a,\mu_b,\mu_c,\mu_x,\mu_y,\mu_z$, and $\mu_w$.

\begin{table}[htdp]
\caption{List of all $7$ different type of columns used in constructing an $(m,3)$-PIR code.}
\begin{center}
\begin{tabular}{|c|c|c|c|c|c|c|c|}
\hline
$h_a$ & $h_b$ & $h_c$ & $h_x$ & $h_y$ & $h_z$ & $h_w$ \\
\hline
$\begin{bmatrix}1 \\ 0 \\ 0 \end{bmatrix}$ &$\begin{bmatrix}0 \\ 1 \\ 0 \end{bmatrix}$ &$\begin{bmatrix}0 \\ 0 \\ 1 \end{bmatrix}$ &$\begin{bmatrix}0 \\ 1 \\ 1 \end{bmatrix}$ &$\begin{bmatrix}1 \\ 0 \\ 1 \end{bmatrix}$ &$\begin{bmatrix}1 \\ 1 \\ 0 \end{bmatrix}$ &$\begin{bmatrix}1 \\ 1 \\ 1 \end{bmatrix}$\\ \hline
\end{tabular}
\end{center}
\label{table:example_s_3}
\end{table}

Since the code has the $k$-server PIR property, there should be $k$ disjoint sets of columns each with $h_a$ as their sum. $h_a$, $h_b+h_z$, $h_c+h_y$, and $h_x+h_w$ are all such possibilities. It is easy to notice that there is no combination of the columns of type $h_b$, $h_c$, and $h_x$ that would give $h_a$. So, each of the $k$ sets should include at least one of the other columns. Therefore,
\begin{align*}
\mu_a+\mu_y+\mu_z+\mu_w\geq k.
\end{align*}
Similar to $\{h_b,h_c,h_x\}$, we have three other sets $\{h_x,h_y,h_z\}$,  $\{h_c,h_z,h_w\}$, and $\{h_b,h_y,h_w\}$ that are incapable of recovering the first data chunk by their own. So we have three more constraints
\begin{align*}
\mu_a+\mu_b+\mu_c+\mu_w\geq k,\\
\mu_a+\mu_b+\mu_x+\mu_y\geq k,\\
\mu_a+\mu_c+\mu_x+\mu_z\geq k.
\end{align*}
Redoing the above argument for the second and the third information chunk, we get the following three \emph{new} constraints
\begin{align*}
\mu_b+\mu_c+\mu_y+\mu_z\geq k,\\
\mu_b+\mu_x+\mu_z+\mu_w\geq k,\\
\mu_c+\mu_x+\mu_y+\mu_w\geq k;
\end{align*}
And, by adding all the above constraints we have
\begin{align*}
A(3,k)=m=\mu_a+\mu_b+\mu_c+\mu_x+\mu_y+\mu_z+\mu_w\geq \frac{7}{4} k.
\end{align*}
It is trivial that when $k$ is divisible by $4$, setting $\mu_a=\mu_b=\mu_c=\mu_x=\mu_y=\mu_z=\mu_w=\frac{k}{4}$ gives the equality. We can indeed use the results from Lemmas~\ref{lem:sublinearity}, and~\ref{lem:even odd} to prove that $A(3,k)=\ceil{\frac{7k}{4}}$.
\end{example}
\begin{IEEEproof}[Proof of Theorem~\ref{thm:small_s}]
For the general $s$, the generator matrix contains at most $2^s-1$ different non-zero columns. Assume $C$ is a $k$-server PIR code with length $m$ and dimension $s$. Therefore, for each $1\leq i \leq s$, one can find $k$ disjoint subsets of the columns with their equal to $$e_i=(\underbrace{0,\cdots,0}_{i-1},1,0, \cdots,0)^t.$$ Similar to the example, we look at all the $(s-1)$-dimensional subspaces $V$ in $\F_2^s$, such that $e_i\notin V$. It is clear that no combination of the columns in $V$ can retrieve $e_i$. So, each of the $k$ subsets should include at least one vector from $V^c$, where $V^c$ denotes the complement of $V$ in $\F_2^s$. Now let $V$ be a subspace of $\F_2^s$ that does not contain the unit vector $e_i$. Then $\sum_{v \in V^c} \mu_v \ge k$ is a constraint involving $\mu_{e_i}$.

There are $$\frac{(2^s-2)(2^s-4)\cdots(2^s-2^{s-2})}{(2^{s-1}-1)(2^{s-1}-2)\cdots(2^{s-1}-2^{s-2})}=2^{s-1}$$ such subspaces for each $i$, which gives us $2^{s-1}$ constraints for each $e_i$. It suffices to show that there are exactly $2^s-1$ unique constraints after merging all these sets. Now we recall that the non-zero codewords of the simplex code of length $2^s-1$ are precisely the supports of all sets of the form $V^c$, where $V$ is an $(s-1)$-dimensional subspace of $\F_2^s$ {(see \cite{Costello} -  page $380$ for proof.)} It is now clear that we have $2^s-1$ unique constraints since there are exactly $2^s-1$ codewords of weight $2^{s-1}$ in the simplex code of length $2^s-1$. Moreover, exactly $2^{s-1}$ of these codewords have value $1$ in a fixed coordinate. In other words, we get the vector $2^{s-1} \one_{2^s-1}$ as the sum of all these codewords. So,
\begin{align*}
&2^{s-1}\sum_{v\in \F_2^s } \mu_v \geq (2^{s}-1)k \;\;\;\;\;\;\;\;\;\;\Longleftrightarrow \\
& A(s,k)= m=\sum_{v\in \F_2^s } \mu_v \geq \frac{(2^{s}-1)}{2^{s-1}}k.
\end{align*}
\end{IEEEproof}

Let us fix $s$. The introduced lower bound in~(\ref{eq:lower_bound_small_s}) along with its equality condition shows that $A(s,k)=O(k)$, when $k$ becomes large. In other words $A(s,k) \sim 2k$ for small $s$ and$k$ large.

\subsection{Storage overhead for fixed $k$}

It was already shown that for fixed $k$, there are elementary constructions to achieve storage overhead $\frac{A(s,k)}{s}$ arbitrary close to $1$. We are yet to determine how fast it decreases. Table~\ref{table:constructions_summary} summarizes the constructions introduced in the previous section with their asymptotic behavior. Note that, the explicit formula for the constructions based on constant-weight codes is only known for $k=3$.

\begin{table}[htdp]
\caption{Comparison of the constructions for $A(s,k)$ with respect to the asymptotic code redundancy ($A(s,k)-s$).}
\begin{center}
\begin{tabular}{|c|c|c|}
\hline
Code construction           & Upper bound on $A(s,k)$                                                                                              & Asymptotic redundancy \\ \hline \hline
Cubic construction            &$A(s,k)\leq s+(k-1)\ceil{s^{\frac{1}{k-1}}}^{k-2}$                                                     & $O(s^{1-\frac{1}{k-1}})$      \\ \hline
Steiner System                 &$A\left(\frac{n(n-1)}{(k-1)(k-2)},k\right)\leq n+\frac{n(n-1)}{(k-1)(k-2)}$                  & $O(s^{\frac{1}{2}})$            \\ \hline
Type-$1$ DTI codes (1)   &$A\left(2^{2 \theta \ell}-(2^{\theta+1}-1)^{\ell},2^{\ell}+2\right)\leq 2^{2\theta \ell} -1$   & $O(s^{\frac{1}{2}})$            \\ \hline
Type-$1$ DTI codes (2)   &$A\left((2^{\lambda}-1)^{\ell}-1,2^{\ell}\right)\leq 2^{\lambda \ell} -1$                   & $O(s^{1-\frac{1}{\ell}})$            \\ \hline
Constant weight codes    &$A\left(\binom{n}{2},3\right)\leq \binom{n}{2}+n$                                                      & $O(s^{\frac{1}{2}})$           \\ \hline
\end{tabular}
\end{center}
\label{table:constructions_summary}
\end{table}

We observe that non of the introduced constructions achieves storage overhead less than $1+O(s^{-\frac{1}{2}})$. However, it is not clear if that is the optimum value one can get. So far, the best (and trivial) lower bound is given by $$A(s,k)\geq s+ O(\log s).$$

\section{PIR Array Codes}\label{sec:Array_PIR}
In all the constructions we presented so far, we assumed that the database was partitioned into $s$ parts, where every server stores $n/s$ bits that were considered to be a single symbol. In this section we seek to extend this idea and let every server store more than a single symbol. For example, we can partition the database into $2s$ parts of $n/(2s)$ bits each such that every server stores two symbols. This can be generalized such that every server stores a fixed number of symbols. One of the benefits of this method to construct PIR codes is that we can support setups in which the number of bits stored in a server is $n/s$ where $s$ is not necessarily an integer. Furthermore, we will show that it is also possible to improve, for some instances of $s$ and $k$, the value of $A(s,k)$ and hence the storage overhead as well. Since every server stores more than a single symbol we treat the code construction as an \emph{array code} and thus we call these codes \emph{PIR array codes}. When a server receives a query $q$ then it resposes with multiple answers corresponding to the number of symbols stored in the server. We illustrate the idea of PIR array codes in the next example. 
\begin{example}\label{ex:PIR array code I}
Assume that the database $\bfx$ is partitioned into 12 parts $\bfx_1,\bfx_2,\ldots,\bfx_{12}$ which are stored in four servers as follows.
\begin{center}
\begin{tabular}{|c|c|c|c|}
\hline
Server 1 & Server 2  & Server 3 & Server 4 \\ \hline \hline
$\bfx_1$ & $\bfx_2$ & $\bfx_3$ & $\bfx_1+\bfx_2+\bfx_3$ \\ \hline
$\bfx_2$ & $\bfx_3$ & $\bfx_1$ &  $\bfx_6$ \\ \hline 
$\bfx_4$ & $\bfx_5$ & $\bfx_4+\bfx_5+\bfx_6$ & $\bfx_4$ \\ \hline
$\bfx_5$ & $\bfx_6$ & $\bfx_8$ & $\bfx_9$  \\ \hline 
$\bfx_7$ & $\bfx_7+\bfx_8+\bfx_9$ & $\bfx_9$ & $\bfx_7$ \\ \hline
$\bfx_8$ & $\bfx_{11}$ & $\bfx_{11}$ &  $\bfx_{12}$ \\ \hline 
$\bfx_{10}+\bfx_{11}+\bfx_{12}$ & $\bfx_{11}$ & $\bfx_{12}$ & $\bfx_{10}$ \\ \hline
\end{tabular}
\end{center}
Thus, every server stores 7 parts, each of $n/12$ bits, so $\frac{n}{12/7}$ bits are stored in each server and the storage overhead is $7/3$. 
Using this code, it is possible to invoke a 3-server linear protocol $\cP(\cQ,\cA,\cC)$. Assume Alice seeks to read the bit $x_{1,i}$ for $i\in[n/12]$, she invokes the algorithm $\cQ$ to receive three queries $\cQ(3,n/12;i) =  (q_1,q_2,q_3)$. The first sever is assigned with the query $q_1$, the second and fourth servers with the query $q_2$ and the third server with the query $q_3$. Each server responds with 7 answers corresponding to the 7 parts it stores. Alice receives all 28 answers but only needs 5 answers to retrieve the value of $x_{1,i}$. From the first server she receives the answer $a_1= \cA(3,1,\bfx_1,q_1)$, from the second server she receives two answers $a_2'= \cA(3,2,\bfx_2,q_2)$ and $a_2'' = \cA(3,2,\bfx_3,q_2)$, from the third server $a_3= \cA(3,3,\bfx_3,q_3)$, and lastly from the fourth server she receives $a_4= \cA(3,2,\bfx_1+\bfx_2+\bfx_3,q_2)$. Note that from the linearity of the protocol $\cP$, we have
$$a_2'+a_2'' + a_4 = \cA(3,2,\bfx_2,q_2) + \cA(3,2,\bfx_3,q_2) + \cA(3,2,\bfx_1+\bfx_2+\bfx_3,q_2) = \cA(3,2,\bfx_1,q_2), $$
and thus $x_{1,i}$ is retrieved by applying the algorithm $\cC$
$$x_{1,i} = \cC(3,n/12;i,a_1,a_2'+a_2'' + a_4,a_3).$$
\end{example}

In the last example, we see that we repeated the same code four times. That was done in order to guarantee that the number of symbols stored in each server is the same. We could instead show only the first two rows of the first code and then claim that by interleaving of the column which stores only one symbol it is possible to guarantee that each server stores the same number of symbols. While we saw that in this example it is possible to construct PIR codes with more flexible parameters, the download communication was increased and we needed only 5 out of the 28 received answers. However, since the number of symbols in each server is fixed (and will be in the constructions in this section) the communication complexity order is not changed.

In general, we refer to an $m_1\times m_2$ array code as a scheme to encode $s$ information bits $x_1,\ldots, x_s$ into an array of size $m_1 \times m_2$. An $(m_1\times m_2,s)$-server coded PIR protocol is defined in a similar way to Definition~\ref{def:coded_PIR}. 
We formally define PIR array codes. 
\begin{definition}\label{def:PIR array code}
A binary $[m_1\times m_2,s]$ linear code will be called a \textbf{$k$-server PIR array code} if for every information bit $x_i$, $i\in [s]$, there exist $k$ mutually disjoint sets $R_{i,1},\ldots,R_{i,k}\subseteq [m_2]$ such that for all $ j\in [k]$, $x_i$ is a linear function of the bits stored in the columns of the set $R_{i,j}$.
\end{definition}

Very similarly to Theorem~\ref{thm:coded_PIR} we conclude that if there exists an $[m_1\times m_2,s]$ $k$-server PIR array code and a $k$-server linear PIR protocol $\cP$ then there exists an $(m_1\times m_2,s)$-server coded PIR protocol that can emulate the protocol $\cP$. Next, we give another example of PIR array code which explicitly improves the storage overhead.
\begin{example}\label{ex:PIR array code II}
We give here a construction of $[2\times 25, 6]$ $15$-server PIR array code. The 6 information bits are denoted by $x_1,x_2,x_3,x_4,x_5,x_6$ and are stored in a $2\times 25$ array as follows:
\begin{center}
\begin{small}
\begin{tabular}{|c|c|c|c|c|c|c|c|c|c|c|c|c|c|c|}
\hline
$1$ & $2$ & $3$ & $4$ & $5$ & $6$ & $7$  & $8$ & $9$ & $10$ & $11$ & $12$ & $13$ & $14$ & $15$ \\ \hline \hline
$x_1$ & $x_1$ & $x_1$ & $x_1$ & $x_1$ & $x_2$ & $x_2$  & $x_2$ & $x_2$ & $x_3$ & $x_3$ & $x_3$ & $x_4$ & $x_4$ & $x_5$ \\ \hline
$x_2$ & $x_3$ & $x_4$ & $x_5$ & $x_6$ & $x_3$ & $x_4$  & $x_5$ & $x_6$ & $x_4$ & $x_5$ & $x_6$ & $x_5$ & $x_6$ & $x_6$ \\ \hline
\end{tabular}

\vspace{2ex}
\begin{tabular}{|c|c|c|c|c|c|c|c|c|c|}
\hline
$16$ & $17$ & $18$ & $19$ & $20$ & $21$ & $22$  & $23$ & $24$ & $25$  \\ \hline \hline
$x_1\hs{+}x_2\hs{+}x_3\hspace{-0.8ex}$  & \hs{$x_1\hs{+}x_2\hs{+}x_4$} & \hs{$x_1\hs{+}x_2\hs{+}x_5$} & \hs{$x_1\hs{+}x_2\hs{+}x_6$} & \hs{$x_1\hs{+}x_3\hs{+}x_4$} & \hs{$x_1\hs{+}x_3\hs{+}x_5$} & \hs{$x_1\hs{+}x_3\hs{+}x_6$}  & \hs{$x_1\hs{+}x_4\hs{+}x_5$} & \hs{$x_1\hs{+}x_4\hs{+}x_6$} & \hs{$x_1\hs{+}x_5\hs{+}x_6$}  \\ \hline
$x_3\hs{+}x_4\hs{+}x_5\hspace{-0.8ex}$  & \hs{$x_3\hs{+}x_5\hs{+}x_6$} & \hs{$x_3\hs{+}x_4\hs{+}x_6$} & \hs{$x_3\hs{+}x_4\hs{+}x_5$} & \hs{$x_2\hs{+}x_5\hs{+}x_6$} & \hs{$x_2\hs{+}x_4\hs{+}x_6$} & \hs{$x_2\hs{+}x_4\hs{+}x_5$}  & \hs{$x_2\hs{+}x_3\hs{+}x_6$} & \hs{$x_2\hs{+}x_3\hs{+}x_5$} & \hs{$x_2\hs{+}x_3\hs{+}x_4$}  \\ \hline
\end{tabular}
\end{small}
\end{center}
The first row specifies the server number. The other two rows indicate the bits which are stored in each column. It is possible to verify that this construction provides a 15-server PIR array code. For example for the first bit we get the following 15 sets:
{\small
$$\{1\}, \{2\}, \{3\}, \{4\}, \{5\}, \{6,16\}, \{7,17\}, \{8,18\}, \{9,19\}, \{10,20\}, \{11,21\}, \{12,22\}, \{13,23\}, \{14,24\}, \{15,25\}
$$}
where it is possible to retrieve the value of $x_1$ by the bits stored in the columns of each group.

The number of bits stored in each server of this example is $n/3$ and thus $s=3$. If we had to use the best construction of an $[m,3]$ 15-server PIR code, then $A(3,15)=26$ servers are required, while here we used only 25 servers. Hence, we managed to improve the storage overhead for $s=3$ and $k=15$.
\end{example}

We extend Example~\ref{ex:PIR array code II} to a general construction of PIR array code. Let $t$ be a fixed integer $t\geq 2$. The number of information bits is $s=t(t+1)$, the number of rows is $m_1 = t$ and the number of columns is $m_2 =  m_2' + m_2''$, where $m_2'= \binom{t(t+1)}{t}$ and $m_2''= \binom{t(t+1)}{t+1}/t$. In the first $m_2'$ columns we simply store all tuples of $t$ bits out of the $t(t+1)$ information bits. In the last $m_2''$ columns we store all possible summations of $t+1$ bits. There are $\binom{t(t+1)}{t+1}$ such summations and since there are $t$ rows, $t$ summations are stored in each column, so the number of columns for this part is $m_2''= \binom{t(t+1)}{t+1}/t$. We also require that in the last $m_2''$ columns every bit appears in exactly one summation. Note that Example~\ref{ex:PIR array code II} is a special case of this construction for $t=2$. A code generated by this construction will be denoted by $\cC_{A-PIR}(t)$. 
\begin{theorem}\label{th:PIR array code}
For any integer $t\geq 2$, the code $\cC_{A-PIR}(t)$ is an $[m_1\times m_2, t(t+1)]$ $k$-server PIR array code where
$$s = t+1, m_1 = t, m_2 =  \binom{t(t+1)}{t} + \frac{\binom{t(t+1)}{t+1}}{t}, k = \binom{t(t+1)}{t},$$
and its storage overhead is 
$$\frac{\binom{t(t+1)}{t} + \binom{t(t+1)}{t+1}/t}{t+1}.$$
\end{theorem}

Table~\ref{table:PIR_array_comparison} compares the improvement in the number of servers, and thus storage overhead, when using the PIR array code $\cC_{A-PIR}(t)$. For $t=2$ and $s=3,k=15$ we know the exact value of $A(s,k)$ according to Table~\ref{table:small values}, and for all other values of $t$ we get a lower bound on the value of $A(s,k)$ according to Theorem~\ref{eq:lower_bound_small_s}.
\begin{table}[htdp]
\caption{Comparison between the code $\cC_{A-PIR}(t)$ and the corresponding best values of $A(s,k)$.}
\begin{center}
\begin{tabular}{|c||c|c|c|c|}
\hline
$t$ & $s$ & $k$ &  $m_2$ & $A(s,k)$  \\ \hline \hline
2 & 3 & 15 & 25 & 26 \\ \hline
3 & 4 & 220 & 385 & $\geq 413$ \\ \hline
4 & 5 & 4845 & 8721 & $\geq 9387$ \\ \hline
5 & 6 & 142506 & 261261 & $\geq 280559$ \\ \hline
\end{tabular}
\end{center}
\label{table:PIR_array_comparison}
\end{table}

The constructions presented in this section are examples for improvements either in the storage overhead or the existence of codes with other parameters which cannot be achieved by the non-array PIR codes. We hope that more constructions will appear to further improve these parameters.

\section{Alternative Contructions}\label{app:alter_cons}
In this section we discuss several more constructions of coded PIR schemes with special properties. First we start with the extension of  binary coded PIR schemes to non-binary codes. Then, we show how other extensions of PIR schemes, namely robust PIR and coalitions PIR, can be adjusted for the coded PIR setup. 

\subsection{Non-binary Coded PIR Schemes}
In this section we extend the results from Section~\ref{sec:cons} to the non-binary setup. Since the construction from Theorem~\ref{thm:coded_PIR} consists of a $k$-server linear PIR protocol and a $k$-server PIR code we require the protocol and code to be over the same field $GF(q)$, where $q$ will be a power of a prime number.

In the extension of Definitions~\ref{def:PIR} and~\ref{def:coded_PIR} we require that the database and the responses of algorithm $\cA$ in the protocol are over the same field $GF(q)$. Therefore, in Definition~\ref{ded:linear_PIR} we also require the linearity of $\cA$ to be over $GF(q)$.
The Definition of $k$-server PIR codes remains the same while the linearity of the sets is over $GF(q)$. For any $s$ and $k$ we denote by $A(s,k)_q$ to be the smallest $m$ such that an $[m,s]$ $k$-server PIR code exists over the field $GF(q)$. The construction of $k$-server PIR protocol in Theorem~\ref{thm:coded_PIR} remains almost identical.

We summarize here the required modifications in this proof, under the assumptions mentioned above.
\begin{enumerate}
\item The database $\bfx$ is partitioned into $s$ parts $\bfx_1,\ldots,\bfx_s$ which are encoded using a generator matrix $G$ over $GF(q)$ as before to receive the coded data which is stored in the $m$ servers $(\bfc_1,\ldots,\bfc_m)$.
\item Alice wants to read the symbol $x_{\ell,i}$. She invokes the algorithm $\cQ$ and receives the $k$ queries as $(q_1,\ldots,q_k)$.
\item We assume that there exist $k$ mutually disjoint sets $R_{\ell,1},\ldots,R_{\ell,k}\subseteq [m]$, such that for $j\in[k]$, we can write $$\bfx_\ell = \sum_{h\in R_{\ell,j}}\alpha_h\bfc_h,$$
    where the coefficients $\alpha_h$ are over the field $GF(q)$.
\item The output of the algorithm $\cQ^*(m,s,n;i)$ is assigned as before with the queries and the received answers $a_h^*$ and $a_h$.
\item For $j\in[k]$ the value of $a_j'$ is calculated according to
    \begin{align*}
 \hspace{-2ex} a_j' & = \sum_{h\in R_{\ell,j}} \hspace{-1ex}\alpha_ha_h = \sum_{h\in R_{\ell,j}} \hspace{-1ex}\alpha_h\cA(k,j,\bfc_h,q_h^*)  \\
    & = \cA(k,j,\hspace{-1ex}\sum_{h\in R_{\ell,j}}\hspace{-1ex}\alpha_h\bfc_h,q_h^*) \hspace{-0.5ex}=\hspace{-0.5ex} \cA(k,j,\bfx_\ell,q_h^*) \hspace{-0.5ex}=\hspace{-0.5ex}  \cA(k,j,\bfx_\ell,q_j).
    \end{align*}
\item Alice calculates the symbol $x_{\ell,i}$ as before according to
    \begin{align*}
    & \cC(k,n/s;i,a_1',a_2',\ldots,a_k') & \\
    & = \cC(k,n/s;i,\cA(k,1,\bfx_\ell,q_j),\ldots,\cA(k,k,\bfx_\ell,q_j)) = x_{\ell,i}. &
    \end{align*}
\end{enumerate}

We can always use the binary constructions as $k$-server PIR codes (assuming for example that the code is given by  a parity check matrix), so we can conclude that $A(s,k)_q\leq A(s,k)$. However, we note that the definition of $k$-server PIR codes is very much related to the recently well-studied \emph{locally recoverable (LRC) codes}~\cite{GHSY12, RPDV14 ,TB14}. A code $\cC$ over $GF(q)$ is said to have \emph{locality} $r$ if every symbol in each codeword from $\cC$ can be recovered by a subset $R$ of at most $r$ other symbols from the codeword. The set $R$ is called a \emph{recovering set} of the symbol. A code $\cC$ is called an \emph{LRC code with locality $r$ and availability $t$} if every symbol has $t$ pairwise disjoint recovering sets, each of size at most $r$. In case the code is systematic while the locality and availability requirements are enforced only on the information symbols then it is called an \emph{LRC code with information locality $r$ and availability $t$}~\cite{RPDV14}. Our definition of $k$-server PIR code is closer to LRC codes with information locality, however we don't require the code to be systematic. Furthermore, the major difference is that we don't restrict the size of the recovering sets. The connection between $k$-server PIR codes and LRC codes with availability is stated as follows. The proof is omitted since it is straightforward.

\begin{theorem}\label{th:LRC_PIR}
If a code $\cC$ is an LRC code with information locality $r$ (or locality $r$) and availability $t=k-1$ then it is a $k$-server PIR code.
\end{theorem}

For the non-binary setup, there are several constructions of LRC codes with availability, see for example~\cite{HYUS15,PHO13,RPDV14,TB14}. While it is not necessarily immediate to find examples where we get better results, in terms of the value of $A_q(s,k)$, than the binary case, it is still possible to improve the minimum distance of the code.

The following example will example demonstrates this idea.
\begin{example}
Assume that $s=2$ and $k=3$, we already sas that $A(2,3)=5$, however the minimum distance of such a code is 3, which is optimal. Let us consider the case $q=4$, then the two information symbols $x_1,x_2$ are encoded to the following five symbols:
$$(x_1,x_2,x_1+x_2,x_1+\alpha x_2,x_1+\alpha^2x_2),$$ where $\alpha$ is a primitive element in $GF(4)$.
It is possible to verify that this is a 3-server PIR code, and its minimum distance is 4, where in the binary case we could only have minimum distance 3.
\end{example}

\subsection{Robust PIR and $t$-private PIR}
Lastly, we briefly note here that our constructions of $k$-server PIR codes can be used also to construct coded PIR protocols for \emph{robust PIR} and \emph{$t$-private PIR}~\cite{WY05}. 

A \emph{$k$-out-of$-\ell$ PIR protocol} is a PIR protocol with the additional property that Alice can compute the value of $x_i$ even though she received only $k$ out of the $\ell$ answers. In order to emulate such a protocol $\cP$ we simply use an $[m,s]$ $\ell$-server PIR code and repeat the same steps as in Theorem~\ref{thm:coded_PIR}. Then, we can emulate the protocol $\cP$ and if at most $\ell-k$ answers were not received, then Alice will still be able to privately recover the value of the bit $x_i$. 

A \emph{$t$-private PIR protocol} is a PIR protocol where every collusion of up to $t$ servers learns no information on the bit Alice seeks to read from the database. Given a $t$-private PIR protocol $\cP$, we follow again the same steps of Theorem~\ref{thm:coded_PIR} to construct an $(m,s)$-server coded PIR protocol $\cP^*$. Since the protocol $\cP$ is $t$-private, we get also that every collusion of $t$ servers learns no information on $i$, the bit that Alice attempts to read. This property results from observing that every $t$ servers have together at most $t$ out of the queries that Alice sends to the servers, and according to the $t$-privacy property of the protocol $\cP$, the same privacy is preserved for the protocol $\cP^*$ as well.

\section{Conclusions and Open Problems}\label{sec:conc} 
A new framework to utilize private information retrieval in distributed storage systems is introduced in this paper. The new scheme is based on the idea of using coding instead of the replications in the traditional PIR protocols, when the storage size of each server is much less than the size of the database. We have shown that among the three main parameters in measuring the quality of $k$-server PIR protocols \emph{i.e.} \emph{communication complexity, computation complexity,} and \emph{storage overhead}, the first two remain the same and the latter improves significantly in the asymptotic regime. In particular, for a fixed $k$ and a limited server size, the storage overhead becomes $1+o(1)$ as the number of servers becomes large.

The optimal storage overhead with the coded PIR is also studied and the explicit value is derived for many cases. 
The presented constructions lead to coded PIR schemes with storage overhead $1+O(s^{-1/2})$ for any fixed $k$, where $s$ is the ratio between the size of the database and the storage size of each server. Hence, it will be interesting to determine whether this asymptotic behavior can be improved. Another research direction is the construction of other coded schemes which are compatible with existing PIR protocols, such as the ones given in Sections~\ref{sec:Array_PIR} and~\ref{app:alter_cons}.

\section*{acknowledgement}
The authors thank Eyal Kushilevitz and Itzhak Tamo for helpful discussions.


\end{document}